\documentclass[letter,traditabstract]{aa}     
\usepackage{graphicx}
\usepackage{txfonts}
\usepackage{xcolor}
\usepackage{booktabs}
\usepackage{subcaption} 
\usepackage{lscape} 
\usepackage{placeins}

\begin{document}

   \title{A closer look at the WISPIT~2 host star}

   \subtitle{Evidence of a spectroscopic binary}

   \author{C. J. B\"{u}rgy\inst{1, 2}\corrauth{buergy@mpia.de}       
   \and M. Benisty\inst{1} 
   \and H. Alqubelat\inst{3}
   \and C.F. Manara\inst{3}
   \and S. Facchini\inst{4}
        }

   \institute{
   Max-Planck Institute for Astronomy (MPIA), K\"{o}nigstuhl 17, 69117 Heidelberg, Germany\\
   \email{buergy@mpia.de}
    \and
    Department of Physics and Astronomy, Heidelberg University, Im Neuenheimer Feld 226, 69120 Heidelberg, Germany
    \and
    European Southern Observatory, Karl-Schwarzschild-Straße 2, 85748 Garching bei M\"{u}nchen, Germany
    \and 
    Dipartimento di Fisica, Universit\`a degli Studi di Milano, Via Celoria 16, 20133 Milano, Italy
    }

   \date{Accepted August 4, 2026}
    
  \abstract
  {While hundreds of protoplanetary discs have been studied in great detail, the detection of protoplanets that are still embedded in their native discs remains rare. WISPIT~2 is only the second laboratory allowing us to directly study planet formation while in progress. The recently discovered system hosts two giant protoplanets in a multi-ringed disc.} 
  {We characterise the WISPIT~2 host star spectroscopically to determine its stellar properties, accretion rate, and inner disc diagnostics to provide a more complete picture of the system.}
  {We present optical and near-infrared spectroscopic observations obtained with the ESO VLT/X-shooter and 2.2\,m/FEROS instruments. We modelled the stellar spectrum to determine the spectral type and effective temperature, analysed the emission lines to estimate the accretion rate, and searched for evidence of a close stellar companion using radial velocity measurements.}
  {Our observations reveal that WISPIT~2 is a spectroscopic binary. The binary has a period of ($4.8\pm~0.1$)~days, which corresponds to a semi-major axis of $0.072$~au or $15.54 R_\odot$, assuming co-planarity with the disc and a circular orbit. The binary system consists of a $\sim$ 0.97\,$M_{\odot}$ primary of spectral type K3 ($T_{\rm eff} \sim 4700$~K) and a $\sim 0.33M_\odot$ secondary (mass ratio $\sim$0.34). We detect weak H$\alpha$ emission below the typical chromospheric level, indicating little to no ongoing accretion onto the young stars, consistent with an accretion rate of $\lesssim 2 \times 10^{-11}\,M_{\odot}\,\mathrm{yr}^{-1}$.}
  {This discovery makes the WISPIT~2 disc the first circumbinary system with directly imaged protoplanets and establishes this system as a unique benchmark for studying planet formation and disc evolution around binary stars.}


    \maketitle
    \nolinenumbers

\section{Introduction}
Thousands of mature exoplanets have now been detected and revealed a remarkable diversity in planetary architectures and properties that likely reflects the wide range of physical conditions present during their formation. To understand this diversity, it is fundamental to study forming protoplanets as they actively shape their natal environments. The PDS~70 system \citep[e.g.,][]{PDS70.Keppler.2018} and the recently discovered WISPIT~2 system \citep{WISPIT2.Close.2025, WISPIT2.vanCapelleveen.2025, WISPIT2.Lawlor.2026} represent two confirmed cases of directly imaged protoplanetary systems. 

This Letter focuses on the latter system. WISPIT~2 (TYC~5709-354-1) is a young ($\sim$5\,Myr) T Tauri system located at a distance of $\sim$133\,pc \citep{WISPIT2.vanCapelleveen.2025}. It hosts two giant planets, WISPIT~2b and WISPIT~2c, with estimated masses of $\sim$4.9 and 8-10 $M_{\rm Jup}$, orbiting at projected separations of $\sim$57\,au and 15\,au, respectively \citep{WISPIT2.vanCapelleveen.2025, WISPIT2.Lawlor.2026}. H$\alpha$ observations indicate that WISPIT~2b is actively accreting from the surrounding circumplanetary material at a rate of approximately $2\times10^{-12}\,M_\odot\,{\rm yr}^{-1}$ \citep{WISPIT2.Close.2025}. In contrast, no H$\alpha$ emission has been detected from the inner planet, WISPIT~2c, suggesting that WISPIT~2b starves it of disc material. Consistent with this scenario,  \citet{Swastik.2026.WISPIT2} recently derived an upper limit in accretion onto the host star of $3.6 \times 10^{-11}\,M_\odot$ yr$^{-1}$. Nevertheless, the host star remains poorly characterised, preventing us from achieving a full picture of the system.  

In this Letter, we present spectroscopic observations of WISPIT~2 that reveal it to be a spectroscopic binary system and provide a detailed spectral characterisation of the primary star, and we study accretion and wind tracers. Sect.~\ref{sec:obs} describes the observations, data reduction, and line analysis. The radial velocity (RV) analysis is presented in Sect.~\ref{sec:RV} and discussed in Sect.~\ref{sec:discussion}.

\section{Spectroscopic observations}
\label{sec:obs}
We obtained two spectra with the X-shooter instrument at the Very Large Telescope (VLT) on April 16 and May 23, 2026 \citep[][DDT 116.2ASZ; PI B\"{u}rgy]{XSHOOTER.Vernet2011}, and five spectra with the fiber-fed extended range optical spectrograph \citep[FEROS;][]{FEROS.Kaufer1999} mounted at the 2.2m MPG/ESO 2.2m telescope between July 9 and July 13, 2026 (DDT 117.2AS7.001; PI Benisty). Details on the observations and data reduction are given in Appendix\, \ref{app:obs}.  

\begin{figure}
    \centering
    \includegraphics[width=1\linewidth]{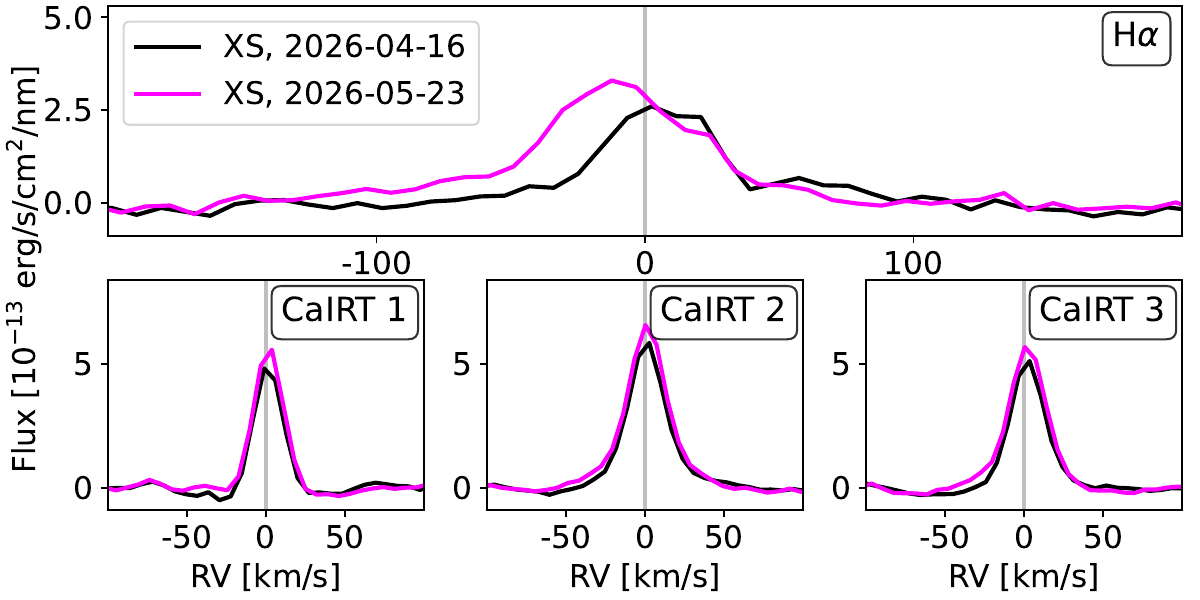}
    \vspace{-5pt}
    \caption{Zoom-in on the H$\alpha$ and CaIRT lines. } 
    \label{fig:ZoomLines}
\end{figure}

The WISPIT~2 X-shooter spectrum shows very few emission lines, which are analysed in more detail below (Fig.\,\ref{fig:ZoomLines} and below), and it is dominated by a stellar photospheric continuum (Fig.\,\ref{fig:XSSpec}). The spectrum shows a clear Li I absorption feature at 670.8 nm, confirming the young nature of the star. To characterise the spectrum, we used the code \texttt{FRAPPE}\footnote{\url{https://github.com/RikClaes/FRAPPE.git}}  \citep[FitteR for Accretion ProPErties of T Tauri stars; ][]{Claes.FRAPPE.2024}, which performs a fit to the entire spectrum using an interpolated grid of empirical class III spectral templates. The best-fit spectral type is K3, with $T_{\rm eff} = 4600$~K, $L_* = 0.63\,L_\odot$, $M_* = 1.1\,M_\odot$, and an age of $\sim$10 Myr at a low extinction of $A_V = 0.15$. This is consistent between the two epochs. FRAPPE was not able to constrain the stellar accretion because there is no UV excess beyond the chromospheric level.

To analyse the emission lines further, we fitted a photospheric model using the grid of PHOENIX high-resolution stellar spectra \citep{PHOENIX.Husser.2013}. Assuming solar metallicity ([Fe/H] = 0.0, [$\alpha$/M] = 0.0), we varied the stellar effective temperature $T_{\rm eff}$ and surface gravity $\log{g}$. The template was convolved with the instrumental resolution of the respective X-shooter arm. Additionally, the model accounted for rotational broadening through the projected stellar rotational velocity $v\sin{i}$ and for continuum veiling through the veiling factor $r = F_{\rm veil} / F_{\rm phot, cont}$. The observed X-shooter spectrum was allowed to be shifted in RV to match to the photospheric spectrum. After normalising the observed and modelled spectrum, we calculated the residuals between the two for a set of spectral regions suitable for tracing photospheric characteristics (see Table \ref{tab:SpecWindows}). For each photospheric model ($T_{\rm eff}$, $\log{g}$), we determined the best-fit RV shift, stellar rotation $v\sin{i}$, and veiling $r$ by minimising the residuals through a least-squares minimisation as implemented in the Python package \texttt{lmfit}. Models at different $T_{\rm eff}$ and $\log{g}$ were then compared through their respective best-fit squared residual sum. We recovered a good fit to the observed photospheric lines at effective temperatures of $T_{\rm eff} = 4500-4800$\,K and surface gravity values of $\log{g} = 4.0-4.5$, with a best-fitting model at $T_{\rm eff} = 4700$\,K, $\log{g} = 4.0$. The best-fit RV shift is $v_{r} = -18.86\pm 0.03$\,km s$^{-1}$ for the first epoch (2026-04-16) and $v_{r} = +25.54\pm 0.03$\,km s$^{-1}$ for the second epoch (2026-05-23). In both epochs, the stellar rotation is consistently constrained to low values ($v\sin{i} \lesssim 15$\,km s$^{-1}$) that are unresolved at the spectral resolution of  X-shooter. We did not detect any significant veiling ($r \sim 0$). The best-fit PHOENIX model and the two X-shooter epochs are shown in Figure \ref{fig:PhotFit}.

We detected H$\alpha$ emission and narrow emission of the calcium infrared triplet (CaIRT) in both epochs. Figure \ref{fig:ZoomLines} shows a closer look at these detected lines after photospheric absorption was removed, and it compares the two epochs.  The H$\alpha$ emission line flux is higher by about 50\% in the second than the first epoch. While the red wing agrees well between the two epochs, the second epoch has a significantly stronger blue wing, which therefore also broadens the line overall. The line parameters are reported in Table \ref{tab:LineDets}. Assuming that the H$\alpha$ traces accretion, we followed the empirical relation derived in \citet{Fiorellino.LaccRelations.2025}; the measured H$\alpha$ flux corresponds to accretion luminosities of $\log{L_{\rm acc, 1}/L_\odot} = -3.63$ and $\log{L_{\rm acc, 2}/L_\odot} = -3.39$. At an approximate mass of $M_* = 1.1\, M_\odot$ and in a radius range of $R_* = 1.5 - 2.0\,R_\odot$, the mass accretion rate of WISPIT~2 lies in a range of $\dot{M}_{\rm acc} = 1.4 \times 10^{-11} - 1.8 \times 10^{-11}$~M$_\odot$~yr$^{-1}$ for the first epoch and $\dot{M}_{\rm acc} = 2.2 \times 10^{-11} - 3.0 \times 10^{-11}$~M$_\odot$~yr$^{-1}$ for the second epoch. However, the observed accretion luminosity is about an order of magnitude below the typical chromospheric noise level in stars of this spectral type \citep{Manara.Frasca.ChromNoise.2017}. The observed H$\alpha$ emission is therefore consistent with a purely chromospheric origin, suggesting that WISPIT~2 is either very weakly or not accreting. 

The CaIRT is fairly similar in line shape and flux between the two epochs. When we compare the line fluxes reported in Table \ref{tab:LineDets}, the second epoch is stronger in line flux by about 14-30\%.  A gallery of typical tracers of accretion and winds is shown in Figs.\,\ref{fig:LineGallery_ep1} and \ref{fig:LineGallery_ep2} for the two epochs without the photospheric contribution. 

\section{Radial velocity variations and orbital fitting}
\label{sec:RV}

\begin{figure}[b]
    \centering
    \includegraphics[width=0.45\textwidth]{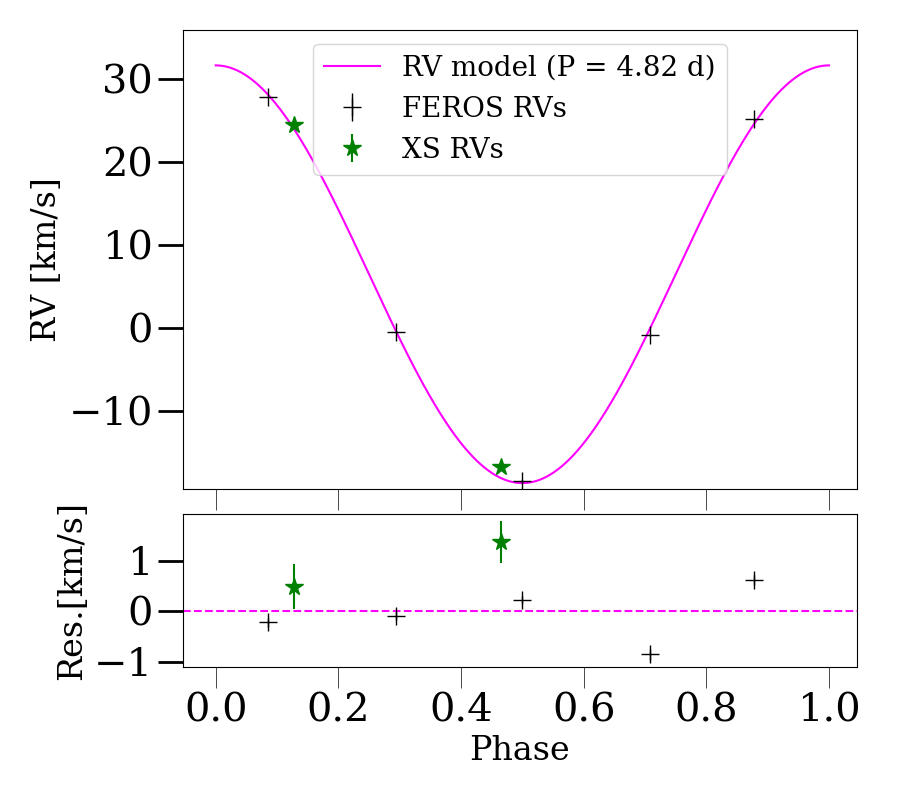}
    \caption{Orbital solution for WISPIT~2 fitted to the RV data (top) with residuals (bottom) as a function of orbital phase. }
    \label{fig:RVCurve}
\end{figure}

Between the two X-shooter epochs, we measured an RV shift of 42.4~km~s$^{-1}$ over 37 days. To investigate the observed RV variation, we targeted WISPIT~2 with five additional epochs using FEROS evenly spaced by $\sim$ 1 day. 

We calculated the RVs using the cross-correlation function (CCF) technique over spectral regions free of strong telluric absorption and emission lines (Tab.~\ref{tab:SpecWindows}). To derive the RV, we cross-correlated each spectrum with the best-fit stellar template from Sect.~\ref{sec:obs}. A Gaussian profile was fitted to each CCF to determine its centroid, which provides the RV measurement, while the uncertainty was taken as the error on the centroid. This procedure was applied consistently to the X-shooter and FEROS data sets. Figure~\ref{fig:CCF} shows the resulting CCFs of WISPIT~2 spanning $\sim$87 days, and the corresponding RV measurements are listed in Tabs.~\ref{tab:XS_ObsLog} and \ref{tab:FEROS_ObsLog}. We modelled the RV measurements using a Markov chain Monte Carlo (MCMC) approach to constrain the orbital parameters of the binary system. Specifically, we fitted a Keplerian model for a single-lined spectroscopic binary (SB1), parametrised by the orbital period ($P$), time of periastron passage ($T_0$), systemic velocity ($\gamma$), and RV semi-amplitude ($K_1$). From these parameters, the model predicts the RV curve, $v_r(t)$, as a function of time $t$ \citep[see][for details]{2026A&A...706A.228A}:

\vspace{-0.45cm}
\begin{equation}
    v_1(t) = V_0 + K_1 \sin\left( \frac{2\pi (t - T_0)}{P} \right)
\end{equation}

We let the \texttt{emcee} sampler run for 200000 steps with an initial burn-in phase of 20000 steps. The resulting corner plots reveal a bimodal posterior distribution, with two possible orbital solutions: one solution with a period of $P\sim4.82$ days (Fig.\,~\ref{fig:RVCurve}), while the other solution yields a period of $P\sim4.62$ days (Fig.\,~\ref{fig:5_15_fit}). Since the former provides a statistically better fit to the RV data, we adopted it as the preferred solution in this work. This period is also consistent with the independent period derived from the TESS photometric periodogram (Fig.\,~\ref{fig:TESS}), providing additional support for this solution. In Table \ref{table:orbital_param}, we report the the orbital parameters of WISPIT~2 derived from the MCMC fit (Fig.\, \ref{fig:MCMC_4_82_days}). We used the parameters corresponding to the maximum likelihood values as best estimates of the model parameters. The uncertainty on the parameters was taken to be half the difference between the 84th and 16th percentiles, providing an estimate of the $1\sigma$ confidence interval. We then calculated the projected mass, separation, and mass function to provide a lower limit on the secondary star mass.

Figure~\ref{fig:HalphaVariab} shows the FEROS H$\alpha$ line profiles, and the equivalent widths measured in all epochs are listed in Table~\ref{tab:HalphaEW}. Despite the low S/N, we find no evidence of significant variations in H$\alpha$ line luminosity over the binary orbital period.

\begin{table}[h]
    \centering
    \caption{Best-fit spectroscopic orbital parameters of WISPIT~2}
    \label{table:orbital_param}
    \renewcommand{\arraystretch}{1.3} 
    \begin{tabular}{lc}
        \toprule
        Element & Value \\
        \midrule
        P (days) & $ 4.8 \pm 0.1 $ \\
        $\gamma$ (km s$^{-1}$)  & $   6.49  \pm 0.065 $ \\
        K (km s$^{-1}$) &   $25.17  \pm 0.05$ \\
        T$_0$ (MJD)  & $ 61101.7967\pm 2.7 $ \\
        \midrule
        $M_1 \sin^3 i$ (M$_\odot$) & $ 0.3523 \pm $0.0002 \\
        f(M) &$0.007\pm0.001$ \\
        $a \sin i$ (au) & $ 0.043 \pm 0.00 $ \\
        $a \sin i$ ($R_{\odot}$) & $ 9.37\pm0.02 $ \\
        \bottomrule
    \end{tabular}
\end{table}

\section{Discussion}
\label{sec:discussion}

\subsection{WISPIT~2 orbital parameters }
The orbital fit of WISPIT~2 reveals RV variations with an amplitude of up to $\pm 25$~km~s$^{-1}$. The consecutively taken epochs with FEROS show rapid RV variations from one day to the next. Our seven RV epochs are consistent with a circular orbit and show small residuals to the fitted model (Fig.\, \ref{fig:RVCurve}). The RV fit and the TESS periodogram (Fig.\,\ref{fig:TESS}) indicate a compatible periodicity, following the binary motion over a period of $\sim 4.82$ days. While a circular configuration provides an excellent fit to the present data, additional RV observations spanning multiple orbital cycles are required to robustly constrain the orbital parameters, including the period, eccentricity $e$, and the orbit orientation $\omega$. 

To derive an estimate of the individual masses of the binary components ($M_{1}$, and $M_{2}$) and their physical separation ($a$), we combined spectroscopic orbital parameters with the total dynamical mass of the system ($M_{\text{tot}}=1.303^{+0.001}_{-0.001}$) estimated from a Keplerian fit to $^{12}$CO ALMA observations (Benisty et al. in prep.). Observational and statistical evidence shows that short-period binaries with $P<20$ days tend to be aligned with the outer disc \citep{2019ApJ...883...22C}. As a result, we adopted the outer disc inclination as derived from dust continuum emission \citep[$i=45.66^\circ$; ][]{WISPIT2.Facchini.2026} as the binary inclination ($i_{\text{bin}}$) to estimate the binary properties. First, we calculated the spectroscopic mass function $f(M)$ from the semi-amplitude of the primary ($K_{1}$) and the orbital period ($P\sim 4.82$ days) for a circular orbit, 
\begin{equation}
    f(M) = \frac{M_{2}^3 \sin^3 i}{(M_{1} + M_{2})^2} = \frac{P K_{1}^3}{2\pi G}
\end{equation}

This allowed us to estimate the primary ($M_{1}=0.97$ $M_{\odot}$) and secondary($M_{2}=0.33$ $M_{\odot}$), the mass ratio ($q=0.34$), and the semi-major axis ($a=0.072$ $\sim 15.54\,R_{\odot}$), which agree with the spectral fit to the X-Shooter observations. For close spectroscopic binaries, \citet{2019ApJ...883...22C} reported that circumbinary discs are typically well aligned with the binary orbital plane, with 68\% of systems having mutual inclinations $\theta < 3^\circ$. For such minor misalignments, the derived mass range can vary in the range of $q = 0.32 - 0.37$, placing the secondary mass in a range of $M_2 = 0.32 - 0.35 M_\odot$. For higher inclinations of the binary orbit, the mass ratio would be lower, and the secondary would have a lower mass.

\subsection{Accretion and disc evolution}
\begin{figure}
    \centering
    \includegraphics[width=0.95\linewidth]{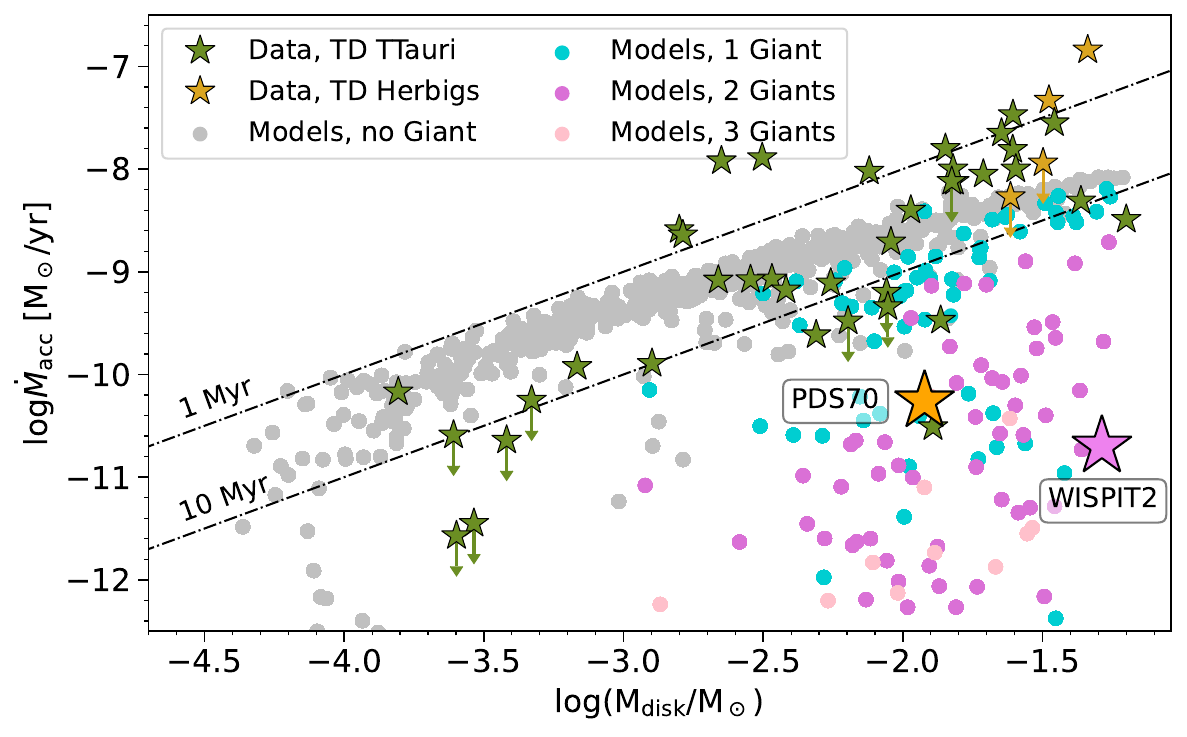}
    \caption{$M_{\rm acc} - M_{\rm disc}$ overview comparing transition disc observations (stars) and model predictions with none, one, two, or three giant planets (dots). The two confirmed giant planet-hosting discs PDS~70 and WISPIT~2 are marked.}
    \label{fig:maccmdisk}
\end{figure}

Our results indicate a very low-mass accretion rate, consistent with the chromospheric noise level. In protoplanetary discs, the relation between stellar accretion and the global properties of protoplanetary discs (e.g. mass and radius) provides an important diagnostic of the physical processes driving disc evolution \citep[e.g.,][]{Lodato.2017.DiscEvolution, Manara2023}. In systems hosting massive planets, however, the accretion flow can also be strongly modified by planet-disc interactions, which regulate the transport of gas through the disc toward the central star. Following \citet{Manara2019}, we compared the observed accretion rates and disc masses with the predictions of planetary population synthesis models for a $\sim$2 Myr old disc population hosting giant planets \citep{Mordasini2009,Mordasini2012} . In Fig.~\ref{fig:maccmdisk} we show these predictions together with observations of transition discs in various star-forming regions \citep{Manara2023,Pinilla2018}. We additionally mark the positions of PDS~70 and WISPIT~2, adopting the dust-continuum-based disc mass estimate as lower limit \citep[][]{WISPIT2.Facchini.2026}. The two systems occupy a similar region in the $\dot{M}_{\rm acc}$–$M_{\rm disc}$ plane, where their low accretion rates are consistent with the expected effect of giant planets that block material beyond their orbits and starve the inner disc regions. The lack of significant H$\alpha$ variability provides further evidence that the transport of material from the outer disc is strongly inhibited by the giant planets, limiting the amount of material that can reach the star. This is consistent with the observed gas- and dust-depleted cavity in $^{12}$CO\,$J=3-2$ within the orbit of WISPIT~2c (Benisty et al. in prep). We note, however, that the binary configuration itself can strongly inhibit accretion onto the stars by dynamically perturbing and truncating the inner disc, thereby reducing the amount of material available to reach inside the orbit of WISPIT~2c.

While the accretion properties of WISPIT~2 are similar to those of PDS~70, the two systems differ in their wind signatures. PDS~70 shows the forbidden [OI] 6300~\AA\ emission line, evidence of a magnetically driven disc wind launched close to the star \citep{CampbellWhite2023,Gaidos2024}. This emission is not detected in WISPIT~2, suggesting that any disc wind is either absent or significantly weaker.

\subsection{Circumbinary discs and planets}
Binary stars are ubiquitous, with $\sim$65\% of stars forming in multiple systems \citep{Duchene.Multiplicity.2013, Offner.Multiplicity}. In circumbinary discs, the central binary clears a cavity that extends about two to three times the binary separation \citep{Cuello.CBD.Review, Artymowicz.1994.CBDCavSize}. This cavity in turn can reduce the efficiency of mass transport onto the stars, potentially allowing circumbinary discs to retain more mass and remain larger at a given age than discs around single stars \citep{Alexander.2012.BinaryAccretion, Ronco.2021.CBDLifeTime}. In this context, WISPIT~2 is particularly reminiscent of V4046~Sgr, a well-studied circumbinary system with a prominent and highly structured disc \citep{Galloway2025} that is in an advanced evolutionary stage ($\sim$20 Myr) compared to typical disc lifetimes \citep{Hernandez2007}. The presence of the binary might be key to the long-lived nature of the disc, potentially providing a more favourable environment for the formation and survival of such massive wide-orbit planets \citep[e.g.,][]{Ruiz2019}.

The population of circumbinary (P-type) planets around mature stars remains small, with only 46 planets known in 37 systems\footnote{\url{https://exoplanet.eu/planets_binary_circum/}}. Most are transit detections and occupy compact configurations, orbiting just beyond the tidally truncated inner edge of the circumbinary disc, which is consistent with inward migration being halted at the cavity edge. Only a handful of circumbinary planets are known beyond 5\,au, and just 2 systems share similarities with WISPIT~2: HD~155555 hosts a 3.8\,M$_{\rm Jup}$ companion at 7.3\,au around a 1.68-day binary \citep{Gratton2024}, and HD~143811 hosts a $\sim$6\,M$_{\rm Jup}$ planet at $\sim$64\,au around an 18.6-day binary \citep{Squicciarini2025}. With its 4.7-day binary and planets at $\sim$15 and $\sim$58\,au, WISPIT~2 belongs to this very small class of wide-orbit planets around compact binaries, but it is so far the only one caught at the time of formation.

\section{Conclusions}
In this Letter, we have shown using VLT/X-Shooter and FEROS spectroscopy that WISPIT~2 is a spectroscopic binary with a period of $4.8\pm~0.1$~days, consisting of a K3 $\sim$0.97\,$M_{\odot}$ primary and a $\sim 0.33M_\odot$ secondary, with little to no ongoing accretion. WISPIT~2 becomes the first system in which two giant protoplanets have been directly imaged within a circumbinary disc, and it provides a key benchmark for studying planet formation and disc evolution around binary stars.

\begin{acknowledgements}
We thank F.~Zagaria for insightful discussions and the anonymous referee for useful feedback. Our Letter is based on observations collected at the European Southern Observatory under ESO programme 116.2ASZ.
MB has received funding from the European Research Council (ERC) under the European Union’s Horizon 2020 research and innovation programme (PROTOPLANETS, grant agreement No. 101002188). HA and CFM are funded by the European Union (ERC, WANDA, 101039452). SF acknowledges financial contribution from the
European Union (ERC, UNVEIL, 101076613). Views and opinions expressed are however those of the authors only and do not necessarily reflect those of the European Union or the European Research Council. Neither the European Union nor the granting authority can be held responsible for them.
      
\end{acknowledgements}

\bibliographystyle{aa} 
\bibliography{bibliography}

@ARTICLE{XSHOOTER.Vernet2011,
       author = {{Vernet}, J. and {Dekker}, H. and {D'Odorico}, S. and {Kaper}, L. and {Kjaergaard}, P. and {Hammer}, F. and {Randich}, S. and {Zerbi}, F. and {Groot}, P.~J. and {Hjorth}, J. and {Guinouard}, I. and {Navarro}, R. and {Adolfse}, T. and {Albers}, P.~W. and {Amans}, J.-P. and {Andersen}, J.~J. and {Andersen}, M.~I. and {Binetruy}, P. and {Bristow}, P. and {Castillo}, R. and {Chemla}, F. and {Christensen}, L. and {Conconi}, P. and {Conzelmann}, R. and {Dam}, J. and {de Caprio}, V. and {de Ugarte Postigo}, A. and {Delabre}, B. and {di Marcantonio}, P. and {Downing}, M. and {Elswijk}, E. and {Finger}, G. and {Fischer}, G. and {Flores}, H. and {Fran{\c{c}}ois}, P. and {Goldoni}, P. and {Guglielmi}, L. and {Haigron}, R. and {Hanenburg}, H. and {Hendriks}, I. and {Horrobin}, M. and {Horville}, D. and {Jessen}, N.~C. and {Kerber}, F. and {Kern}, L. and {Kiekebusch}, M. and {Kleszcz}, P. and {Klougart}, J. and {Kragt}, J. and {Larsen}, H.~H. and {Lizon}, J.-L. and {Lucuix}, C. and {Mainieri}, V. and {Manuputy}, R. and {Martayan}, C. and {Mason}, E. and {Mazzoleni}, R. and {Michaelsen}, N. and {Modigliani}, A. and {Moehler}, S. and {M{\o}ller}, P. and {Norup S{\o}rensen}, A. and {N{\o}rregaard}, P. and {P{\'e}roux}, C. and {Patat}, F. and {Pena}, E. and {Pragt}, J. and {Reinero}, C. and {Rigal}, F. and {Riva}, M. and {Roelfsema}, R. and {Royer}, F. and {Sacco}, G. and {Santin}, P. and {Schoenmaker}, T. and {Spano}, P. and {Sweers}, E. and {Ter Horst}, R. and {Tintori}, M. and {Tromp}, N. and {van Dael}, P. and {van der Vliet}, H. and {Venema}, L. and {Vidali}, M. and {Vinther}, J. and {Vola}, P. and {Winters}, R. and {Wistisen}, D. and {Wulterkens}, G. and {Zacchei}, A.},
        title = "{X-shooter, the new wide band intermediate resolution spectrograph at the ESO Very Large Telescope}",
      journal = {\aap},
         year = 2011,
        month = dec,
       volume = {536},
          eid = {A105},
        pages = {A105},
          doi = {10.1051/0004-6361/201117752},
archivePrefix = {arXiv},
       eprint = {1110.1944},
 primaryClass = {astro-ph.IM},
       adsurl = {https://ui.adsabs.harvard.edu/abs/2011A&A...536A.105V}
}

@ARTICLE{FEROS.Kaufer1999,
       author = {{Kaufer}, A. and {Stahl}, O. and {Tubbesing}, S. and {N{\o}rregaard}, P. and {Avila}, G. and {Francois}, P. and {Pasquini}, L. and {Pizzella}, A.},
        title = "{Commissioning FEROS, the new high-resolution spectrograph at La-Silla.}",
      journal = {The Messenger},
         year = 1999,
        month = mar,
       volume = {95},
        pages = {8-12},
       adsurl = {https://ui.adsabs.harvard.edu/abs/1999Msngr..95....8K}
}

@INPROCEEDINGS{XSHOOTER.Pipeline.Modigliani2010,
       author = {{Modigliani}, Andrea and {Goldoni}, Paolo and {Royer}, Fr{\'e}d{\'e}ric and {Haigron}, Regis and {Guglielmi}, Laurent and {Fran{\c{c}}ois}, Patrick and {Horrobin}, Matthew and {Bristow}, Paul and {Vernet}, Joel and {Moehler}, Sabine and {Kerber}, Florian and {Ballester}, Pascal and {Mason}, Elena and {Christensen}, Lise},
        title = "{The X-shooter pipeline}",
    booktitle = {Observatory Operations: Strategies, Processes, and Systems III},
         year = 2010,
       editor = {{Silva}, David R. and {Peck}, Alison B. and {Soifer}, B. Thomas},
       series = {Society of Photo-Optical Instrumentation Engineers (SPIE) Conference Series},
       volume = {7737},
        month = jul,
          eid = {773728},
        pages = {773728},
          doi = {10.1117/12.857211},
       adsurl = {https://ui.adsabs.harvard.edu/abs/2010SPIE.7737E..28M}
}

@ARTICLE{Ruiz2019,
       author = {{Ru{\'\i}z-Rodr{\'\i}guez}, Dary and {Kastner}, Joel H. and {Dong}, Ruobing and {Principe}, David A. and {Andrews}, Sean M. and {Wilner}, David J.},
        title = "{Constraints on a Putative Planet Sculpting the V4046 Sagittarii Circumbinary Disk}",
      journal = {\aj},
         year = 2019,
        month = jun,
       volume = {157},
       number = {6},
          eid = {237},
        pages = {237},
          doi = {10.3847/1538-3881/ab1c58},
archivePrefix = {arXiv},
       eprint = {1904.09866},
 primaryClass = {astro-ph.EP},
       adsurl = {https://ui.adsabs.harvard.edu/abs/2019AJ....157..237R}
}

@ARTICLE{Squicciarini2025,
       author = {{Squicciarini}, V. and {Mazoyer}, J. and {Wilkinson}, C. and {Lagrange}, A.-M. and {Delorme}, P. and {Radcliffe}, A. and {Flasseur}, O. and {Kiefer}, F. and {Alecian}, E.},
        title = "{GPI+SPHERE detection of a 6.1 M$_{Jup}$ circumbinary planet around HD 143811}",
      journal = {\aap},
         year = 2025,
        month = oct,
       volume = {702},
          eid = {L10},
        pages = {L10},
          doi = {10.1051/0004-6361/202557104},
archivePrefix = {arXiv},
       eprint = {2509.06009},
 primaryClass = {astro-ph.EP},
       adsurl = {https://ui.adsabs.harvard.edu/abs/2025A&A...702L..10S}
}

@ARTICLE{Gratton2024,
       author = {{Gratton}, R. and {Bonavita}, M. and {Mesa}, D. and {Desidera}, S. and {Zurlo}, A. and {Marino}, S. and {D'Orazi}, V. and {Rigliaco}, E. and {Nascimbeni}, V. and {Barbato}, D. and {Columba}, G. and {Squicciarini}, V.},
        title = "{Stellar companions and Jupiter-like planets in young associations}",
      journal = {\aap},
         year = 2024,
        month = may,
       volume = {685},
          eid = {A119},
        pages = {A119},
          doi = {10.1051/0004-6361/202348393},
archivePrefix = {arXiv},
       eprint = {2402.02148},
 primaryClass = {astro-ph.EP},
       adsurl = {https://ui.adsabs.harvard.edu/abs/2024A&A...685A.119G}
}

@article{Hernandez2007,
  author = {{Hern{\'a}ndez}, Jes{\'u}s and {Hartmann}, Lee and {Megeath}, S. T. and others},
  title = {Spitzer Space Telescope Study of Disks in the Young $\sigma$ Orionis Cluster},
  journal = {The Astrophysical Journal},
  volume = {662},
  pages = {1067--1081},
  year = {2007},
  doi = {10.1086/518767}
}

@ARTICLE{Galloway2025,
       author = {{Galloway-Sprietsma}, Maria and {Bae}, Jaehan and {Izquierdo}, Andr{\'e}s F. and {Stadler}, Jochen and {Longarini}, Cristiano and {Teague}, Richard and {Andrews}, Sean M. and {Winter}, Andrew J. and {Benisty}, Myriam and {Facchini}, Stefano and {Rosotti}, Giovanni and {Zawadzki}, Brianna and {Pinte}, Christophe and {Fasano}, Daniele and {Barraza-Alfaro}, Marcelo and {Cataldi}, Gianni and {Cuello}, Nicol{\'a}s and {Curone}, Pietro and {Czekala}, Ian and {Flock}, Mario and {Fukagawa}, Misato and {Gardner}, Charles H. and {Garg}, Himanshi and {Hall}, Cassandra and {Huang}, Jane and {Ilee}, John D. and {Kanagawa}, Kazuhiro and {Lesur}, Geoffroy and {Lodato}, Giuseppe and {Loomis}, Ryan A. and {Menard}, Francois and {Orihara}, Ryuta and {Price}, Daniel J. and {Wafflard-Fernandez}, Gaylor and {Wilner}, David J. and {W{\"o}lfer}, Lisa and {Yen}, Hsi-Wei and {Yoshida}, Tomohiro C.},
        title = "{exoALMA. V. Gaseous Emission Surfaces and Temperature Structures}",
      journal = {\apjl},
         year = 2025,
        month = may,
       volume = {984},
       number = {1},
          eid = {L10},
        pages = {L10},
          doi = {10.3847/2041-8213/adc437},
archivePrefix = {arXiv},
       eprint = {2504.19902},
 primaryClass = {astro-ph.EP},
       adsurl = {https://ui.adsabs.harvard.edu/abs/2025ApJ...984L..10G}
}

@ARTICLE{CampbellWhite2023,
       author = {{Campbell-White}, Justyn and {Manara}, Carlo F. and {Benisty}, Myriam and {Natta}, Antonella and {Claes}, Rik A.~B. and {Frasca}, Antonio and {Bae}, Jaehan and {Facchini}, Stefano and {Isella}, Andrea and {P{\'e}rez}, Laura and {Pinilla}, Paola and {Sicilia-Aguilar}, Aurora and {Teague}, Richard},
        title = "{A Magnetically Driven Disk Wind in the Inner Disk of PDS 70}",
      journal = {\apj},
         year = 2023,
        month = oct,
       volume = {956},
       number = {1},
          eid = {25},
        pages = {25},
          doi = {10.3847/1538-4357/acf0c0},
archivePrefix = {arXiv},
       eprint = {2308.09554},
 primaryClass = {astro-ph.SR},
       adsurl = {https://ui.adsabs.harvard.edu/abs/2023ApJ...956...25C}
}

@ARTICLE{EsoReflex.Freudling2013,
       author = {{Freudling}, W. and {Romaniello}, M. and {Bramich}, D.~M. and {Ballester}, P. and {Forchi}, V. and {Garc{\'\i}a-Dabl{\'o}}, C.~E. and {Moehler}, S. and {Neeser}, M.~J.},
        title = "{Automated data reduction workflows for astronomy. The ESO Reflex environment}",
      journal = {\aap},
         year = 2013,
        month = nov,
       volume = {559},
          eid = {A96},
        pages = {A96},
          doi = {10.1051/0004-6361/201322494},
archivePrefix = {arXiv},
       eprint = {1311.5411},
 primaryClass = {astro-ph.IM},
       adsurl = {https://ui.adsabs.harvard.edu/abs/2013A&A...559A..96F}
}

@ARTICLE{Molecfit.Smette2015,
       author = {{Smette}, A. and {Sana}, H. and {Noll}, S. and {Horst}, H. and {Kausch}, W. and {Kimeswenger}, S. and {Barden}, M. and {Szyszka}, C. and {Jones}, A.~M. and {Gallenne}, A. and {Vinther}, J. and {Ballester}, P. and {Taylor}, J.},
        title = "{Molecfit: A general tool for telluric absorption correction. I. Method and application to ESO instruments}",
      journal = {\aap},
         year = 2015,
        month = apr,
       volume = {576},
          eid = {A77},
        pages = {A77},
          doi = {10.1051/0004-6361/201423932},
archivePrefix = {arXiv},
       eprint = {1501.07239},
 primaryClass = {astro-ph.IM},
       adsurl = {https://ui.adsabs.harvard.edu/abs/2015A&A...576A..77S}
}

@ARTICLE{Molecfit.Kausch2015,
       author = {{Kausch}, W. and {Noll}, S. and {Smette}, A. and {Kimeswenger}, S. and {Barden}, M. and {Szyszka}, C. and {Jones}, A.~M. and {Sana}, H. and {Horst}, H. and {Kerber}, F.},
        title = "{Molecfit: A general tool for telluric absorption correction. II. Quantitative evaluation on ESO-VLT/X-Shooterspectra}",
      journal = {\aap},
         year = 2015,
        month = apr,
       volume = {576},
          eid = {A78},
        pages = {A78},
          doi = {10.1051/0004-6361/201423909},
archivePrefix = {arXiv},
       eprint = {1501.07265},
 primaryClass = {astro-ph.IM},
       adsurl = {https://ui.adsabs.harvard.edu/abs/2015A&A...576A..78K}
}

@ARTICLE{Gaidos2024,
       author = {{Gaidos}, Eric and {Thanathibodee}, Thanawuth and {Hoffman}, Andrew and {Ong}, Joel and {Hinkle}, Jason and {Shappee}, Benjamin J. and {Banzatti}, Andrea},
        title = "{The Dynamic, Chimeric Inner Disk of PDS 70}",
      journal = {\apj},
         year = 2024,
        month = may,
       volume = {966},
       number = {2},
          eid = {167},
        pages = {167},
          doi = {10.3847/1538-4357/ad3447},
archivePrefix = {arXiv},
       eprint = {2403.09970},
 primaryClass = {astro-ph.EP},
       adsurl = {https://ui.adsabs.harvard.edu/abs/2024ApJ...966..167G}
}

@INPROCEEDINGS{Manara2023,
       author = {{Manara}, C.~F. and {Ansdell}, M. and {Rosotti}, G.~P. and {Hughes}, A.~M. and {Armitage}, P.~J. and {Lodato}, G. and {Williams}, J.~P.},
        title = "{Demographics of Young Stars and their Protoplanetary Disks: Lessons Learned on Disk Evolution and its Connection to Planet Formation}",
    booktitle = {Protostars and Planets VII},
         year = 2023,
       editor = {{Inutsuka}, S. and {Aikawa}, Y. and {Muto}, T. and {Tomida}, K. and {Tamura}, M.},
       series = {Astronomical Society of the Pacific Conference Series},
       volume = {534},
        month = jul,
        pages = {539},
          doi = {10.48550/arXiv.2203.09930},
archivePrefix = {arXiv},
       eprint = {2203.09930},
 primaryClass = {astro-ph.SR},
       adsurl = {https://ui.adsabs.harvard.edu/abs/2023ASPC..534..539M}
}

@ARTICLE{Mordasini2009,
       author = {{Mordasini}, C. and {Alibert}, Y. and {Benz}, W.},
        title = "{Extrasolar planet population synthesis. I. Method, formation tracks, and mass-distance distribution}",
      journal = {\aap},
         year = 2009,
        month = jul,
       volume = {501},
       number = {3},
        pages = {1139-1160},
          doi = {10.1051/0004-6361/200810301},
archivePrefix = {arXiv},
       eprint = {0904.2524},
 primaryClass = {astro-ph.EP},
       adsurl = {https://ui.adsabs.harvard.edu/abs/2009A&A...501.1139M}
}

@ARTICLE{Mordasini2012,
       author = {{Mordasini}, C. and {Alibert}, Y. and {Klahr}, H. and {Henning}, T.},
        title = "{Characterization of exoplanets from their formation. I. Models of combined planet formation and evolution}",
      journal = {\aap},
         year = 2012,
        month = nov,
       volume = {547},
          eid = {A111},
        pages = {A111},
          doi = {10.1051/0004-6361/201118457},
archivePrefix = {arXiv},
       eprint = {1206.6103},
 primaryClass = {astro-ph.EP},
       adsurl = {https://ui.adsabs.harvard.edu/abs/2012A&A...547A.111M}
}

@ARTICLE{Manara2019,
       author = {{Manara}, C.~F. and {Mordasini}, C. and {Testi}, L. and {Williams}, J.~P. and {Miotello}, A. and {Lodato}, G. and {Emsenhuber}, A.},
        title = "{Constraining disk evolution prescriptions of planet population synthesis models with observed disk masses and accretion rates}",
      journal = {\aap},
         year = 2019,
        month = nov,
       volume = {631},
          eid = {L2},
        pages = {L2},
          doi = {10.1051/0004-6361/201936488},
archivePrefix = {arXiv},
       eprint = {1909.08485},
 primaryClass = {astro-ph.EP},
       adsurl = {https://ui.adsabs.harvard.edu/abs/2019A&A...631L...2M}
}

@ARTICLE{Pinilla2018,
       author = {{Pinilla}, P. and {Tazzari}, M. and {Pascucci}, I. and {Youdin}, A.~N. and {Garufi}, A. and {Manara}, C.~F. and {Testi}, L. and {van der Plas}, G. and {Barenfeld}, S.~A. and {Canovas}, H. and {Cox}, E.~G. and {Hendler}, N.~P. and {P{\'e}rez}, L.~M. and {van der Marel}, N.},
        title = "{Homogeneous Analysis of the Dust Morphology of Transition Disks Observed with ALMA: Investigating Dust Trapping and the Origin of the Cavities}",
      journal = {\apj},
         year = 2018,
        month = may,
       volume = {859},
       number = {1},
          eid = {32},
        pages = {32},
          doi = {10.3847/1538-4357/aabf94},
archivePrefix = {arXiv},
       eprint = {1804.07301},
 primaryClass = {astro-ph.EP},
       adsurl = {https://ui.adsabs.harvard.edu/abs/2018ApJ...859...32P}
}

@ARTICLE{PDS70.Keppler.2018,
       author = {{Keppler}, M. and {Benisty}, M. and {M{\"u}ller}, A. and {Henning}, Th. and {van Boekel}, R. and {Cantalloube}, F. and {Ginski}, C. and {van Holstein}, R.~G. and {Maire}, A.-L. and {Pohl}, A. and {Samland}, M. and {Avenhaus}, H. and {Baudino}, J.-L. and {Boccaletti}, A. and {de Boer}, J. and {Bonnefoy}, M. and {Chauvin}, G. and {Desidera}, S. and {Langlois}, M. and {Lazzoni}, C. and {Marleau}, G.-D. and {Mordasini}, C. and {Pawellek}, N. and {Stolker}, T. and {Vigan}, A. and {Zurlo}, A. and {Birnstiel}, T. and {Brandner}, W. and {Feldt}, M. and {Flock}, M. and {Girard}, J. and {Gratton}, R. and {Hagelberg}, J. and {Isella}, A. and {Janson}, M. and {Juhasz}, A. and {Kemmer}, J. and {Kral}, Q. and {Lagrange}, A.-M. and {Launhardt}, R. and {Matter}, A. and {M{\'e}nard}, F. and {Milli}, J. and {Molli{\`e}re}, P. and {Olofsson}, J. and {P{\'e}rez}, L. and {Pinilla}, P. and {Pinte}, C. and {Quanz}, S.~P. and {Schmidt}, T. and {Udry}, S. and {Wahhaj}, Z. and {Williams}, J.~P. and {Buenzli}, E. and {Cudel}, M. and {Dominik}, C. and {Galicher}, R. and {Kasper}, M. and {Lannier}, J. and {Mesa}, D. and {Mouillet}, D. and {Peretti}, S. and {Perrot}, C. and {Salter}, G. and {Sissa}, E. and {Wildi}, F. and {Abe}, L. and {Antichi}, J. and {Augereau}, J.-C. and {Baruffolo}, A. and {Baudoz}, P. and {Bazzon}, A. and {Beuzit}, J.-L. and {Blanchard}, P. and {Brems}, S.~S. and {Buey}, T. and {De Caprio}, V. and {Carbillet}, M. and {Carle}, M. and {Cascone}, E. and {Cheetham}, A. and {Claudi}, R. and {Costille}, A. and {Delboulb{\'e}}, A. and {Dohlen}, K. and {Fantinel}, D. and {Feautrier}, P. and {Fusco}, T. and {Giro}, E. and {Gluck}, L. and {Gry}, C. and {Hubin}, N. and {Hugot}, E. and {Jaquet}, M. and {Le Mignant}, D. and {Llored}, M. and {Madec}, F. and {Magnard}, Y. and {Martinez}, P. and {Maurel}, D. and {Meyer}, M. and {M{\"o}ller-Nilsson}, O. and {Moulin}, T. and {Mugnier}, L. and {Orign{\'e}}, A. and {Pavlov}, A. and {Perret}, D. and {Petit}, C. and {Pragt}, J. and {Puget}, P. and {Rabou}, P. and {Ramos}, J. and {Rigal}, F. and {Rochat}, S. and {Roelfsema}, R. and {Rousset}, G. and {Roux}, A. and {Salasnich}, B. and {Sauvage}, J.-F. and {Sevin}, A. and {Soenke}, C. and {Stadler}, E. and {Suarez}, M. and {Turatto}, M. and {Weber}, L.},
        title = "{Discovery of a planetary-mass companion within the gap of the transition disk around PDS 70}",
      journal = {\aap},
         year = 2018,
        month = sep,
       volume = {617},
          eid = {A44},
        pages = {A44},
          doi = {10.1051/0004-6361/201832957},
archivePrefix = {arXiv},
       eprint = {1806.11568},
 primaryClass = {astro-ph.EP},
       adsurl = {https://ui.adsabs.harvard.edu/abs/2018A&A...617A..44K}
}

@ARTICLE{WISPIT2.Close.2025,
       author = {{Close}, Laird M. and {van Capelleveen}, Richelle F. and {Weible}, Gabriel and {Wagner}, Kevin and {Haffert}, Sebastiaan Y. and {Males}, Jared R. and {Ilyin}, Ilya and {Kenworthy}, Matthew A. and {Li}, Jialin and {Long}, Joseph D. and {Ertel}, Steve and {Ginski}, Christian and {Weinberger}, Alycia J. and {Follette}, Kate and {Liberman}, Joshua and {Twitchell}, Katie and {Johnson}, Parker and {Kueny}, Jay and {Apai}, Daniel and {Doyon}, Rene and {Foster}, Warren and {Gasho}, Victor and {Van Gorkom}, Kyle and {Guyon}, Olivier and {Kautz}, Maggie Y. and {McLeod}, Avalon and {McEwen}, Eden and {Pearce}, Logan and {Schatz}, Lauren and {Hedglen}, Alexander D. and {Wu}, Ya-Lin and {Isbell}, Jacob and {Power}, Jenny and {Carlson}, Jared and {Close}, Emmeline and {Tonucci}, Elena and {Mars}, Matthijs},
        title = "{Wide Separation Planets in Time (WISPIT): Discovery of a Gap H{\ensuremath{\alpha}} Protoplanet WISPIT 2b with MagAO-X}",
      journal = {\apjl},
         year = 2025,
        month = sep,
       volume = {990},
       number = {1},
          eid = {L9},
        pages = {L9},
          doi = {10.3847/2041-8213/adf7a5},
archivePrefix = {arXiv},
       eprint = {2508.19046},
 primaryClass = {astro-ph.EP},
       adsurl = {https://ui.adsabs.harvard.edu/abs/2025ApJ...990L...9C}
}

@ARTICLE{WISPIT2.vanCapelleveen.2025,
       author = {{van Capelleveen}, Richelle F. and {Kenworthy}, Matthew A. and {Ginski}, Christian and {Mamajek}, Eric E. and {Bohn}, Alexander J. and {Landman}, Rico and {Stolker}, Tomas and {Zhang}, Yapeng and {van der Marel}, Nienke and {Snellen}, Ignas},
        title = "{WIde Separation Planets In Time (WISPIT): Two directly imaged exoplanets around the Sun-like stellar binary WISPIT 1}",
      journal = {\aap},
         year = 2025,
        month = dec,
       volume = {704},
          eid = {A221},
        pages = {A221},
          doi = {10.1051/0004-6361/202556584},
archivePrefix = {arXiv},
       eprint = {2508.18456},
 primaryClass = {astro-ph.EP},
       adsurl = {https://ui.adsabs.harvard.edu/abs/2025A&A...704A.221V}
}

@ARTICLE{WISPIT2.Lawlor.2026,
       author = {{Lawlor}, Chloe and {van Capelleveen}, Richelle F. and {Bourdarot}, Guillaume and {Ginski}, Christian and {Kenworthy}, Matthew A. and {Stolker}, Tomas and {Close}, Laird and {Bohn}, Alexander J. and {Eisenhauer}, Frank and {Garcia}, Paulo and {H{\"o}nig}, Sebastian F. and {Kammerer}, Jens and {Kreidberg}, Laura and {Lacour}, Sylvestre and {Le Bouquin}, Jean-Baptiste and {Mamajek}, Eric and {Nowak}, Mathias and {Paumard}, Thibaut and {Straubmeier}, Christian and {van der Marel}, Nienke and {The Exogravity Collaboration}},
        title = "{Direct Spectroscopic Confirmation of the Young Embedded Protoplanet WISPIT 2c}",
      journal = {\apjl},
         year = 2026,
        month = apr,
       volume = {1000},
       number = {2},
          eid = {L38},
        pages = {L38},
          doi = {10.3847/2041-8213/ae4b3b},
archivePrefix = {arXiv},
       eprint = {2603.22085},
 primaryClass = {astro-ph.EP},
       adsurl = {https://ui.adsabs.harvard.edu/abs/2026ApJ..1000L..38L}
}

@ARTICLE{WISPIT2.Facchini.2026,
       author = {{Facchini}, Stefano and {Curone}, Pietro and {Benisty}, Myriam and {Zagaria}, Francesco and {Teague}, Richard and {Cugno}, Gabriele and {Bae}, Jaehan},
        title = "{A 2 au Resolution View by ALMA of the Planet-hosting WISPIT 2 Disk}",
      journal = {\apjl},
         year = 2026,
        month = feb,
       volume = {998},
       number = {1},
          eid = {L16},
        pages = {L16},
          doi = {10.3847/2041-8213/ae3c0a},
archivePrefix = {arXiv},
       eprint = {2601.15948},
 primaryClass = {astro-ph.EP},
       adsurl = {https://ui.adsabs.harvard.edu/abs/2026ApJ...998L..16F}
}

@ARTICLE{PHOENIX.Husser.2013,
       author = {{Husser}, T.-O. and {Wende-von Berg}, S. and {Dreizler}, S. and {Homeier}, D. and {Reiners}, A. and {Barman}, T. and {Hauschildt}, P.~H.},
        title = "{A new extensive library of PHOENIX stellar atmospheres and synthetic spectra}",
      journal = {\aap},
         year = 2013,
        month = may,
       volume = {553},
          eid = {A6},
        pages = {A6},
          doi = {10.1051/0004-6361/201219058},
archivePrefix = {arXiv},
       eprint = {1303.5632},
 primaryClass = {astro-ph.SR},
       adsurl = {https://ui.adsabs.harvard.edu/abs/2013A&A...553A...6H}
}

@ARTICLE{Brahm.2017.Ceres.DataRed,
       author = {{Brahm}, Rafael and {Jord{\'a}n}, Andr{\'e}s and {Espinoza}, N{\'e}stor},
        title = "{CERES: A Set of Automated Routines for Echelle Spectra}",
      journal = {\pasp},
         year = 2017,
        month = mar,
       volume = {129},
       number = {973},
        pages = {034002},
          doi = {10.1088/1538-3873/aa5455},
archivePrefix = {arXiv},
       eprint = {1609.02279},
 primaryClass = {astro-ph.IM},
       adsurl = {https://ui.adsabs.harvard.edu/abs/2017PASP..129c4002B}
}

@ARTICLE{2026A&A...706A.228A,
       author = {{Alqubelat}, Hala and {Manara}, Carlo F. and {Campbell-White}, Justyn and {Petr-Gotzens}, Monika G. and {Tofflemire}, Benjamin M. and {Banzatti}, Andrea and {Ragusa}, Enrico and {Whelan}, Emma T. and {Bourdarot}, Guillaume and {Dougados}, Catherine and {Fiorellino}, Eleonora and {Mills}, Sean I.},
        title = "{Coordinated space-and ground-based monitoring of accretion bursts in a protoplanetary disc: The orbital and accretion properties of DQ Tau}",
      journal = {\aap},
         year = 2026,
        month = feb,
       volume = {706},
          eid = {A228},
        pages = {A228},
          doi = {10.1051/0004-6361/202557425},
archivePrefix = {arXiv},
       eprint = {2511.08311},
 primaryClass = {astro-ph.SR},
       adsurl = {https://ui.adsabs.harvard.edu/abs/2026A&A...706A.228A}
}

@ARTICLE{Claes.FRAPPE.2024,
       author = {{Claes}, R.~A.~B. and {Campbell-White}, J. and {Manara}, C.~F. and {Frasca}, A. and {Natta}, A. and {Alcal{\'a}}, J.~M. and {Armeni}, A. and {Fang}, M. and {Lovell}, J.~B. and {Stelzer}, B. and {Venuti}, L. and {Wyatt}, M. and {Queitsch}, A.},
        title = "{FitteR for Accretion ProPErties of T Tauri stars (FRAPPE): A new approach to use class III spectra to derive stellar and accretion properties}",
      journal = {\aap},
         year = 2024,
        month = oct,
       volume = {690},
          eid = {A122},
        pages = {A122},
          doi = {10.1051/0004-6361/202450885},
archivePrefix = {arXiv},
       eprint = {2407.11866},
 primaryClass = {astro-ph.SR},
       adsurl = {https://ui.adsabs.harvard.edu/abs/2024A&A...690A.122C}
}

@ARTICLE{Fiorellino.LaccRelations.2025,
       author = {{Fiorellino}, E. and {Alcal{\'a}}, J.~M. and {Manara}, C.~F. and {Pittman}, C.~V. and {{\'A}brah{\'a}m}, P. and {Venuti}, L. and {Cabrit}, S. and {Claes}, R. and {Fang}, M. and {K{\'o}sp{\'a}l}, {\'A}. and {Lodato}, G. and {Mauco}, K. and {Tychoniec}, {\L}.},
        title = "{PENELLOPE: VII. Revisiting empirical relations to measure accretion luminosity}",
      journal = {\aap},
         year = 2025,
        month = dec,
       volume = {704},
          eid = {A42},
        pages = {A42},
          doi = {10.1051/0004-6361/202556603},
archivePrefix = {arXiv},
       eprint = {2509.21078},
 primaryClass = {astro-ph.SR},
       adsurl = {https://ui.adsabs.harvard.edu/abs/2025A&A...704A..42F}
}

@ARTICLE{2019ApJ...883...22C,
       author = {{Czekala}, Ian and {Chiang}, Eugene and {Andrews}, Sean M. and {Jensen}, Eric L.~N. and {Torres}, Guillermo and {Wilner}, David J. and {Stassun}, Keivan G. and {Macintosh}, Bruce},
        title = "{The Degree of Alignment between Circumbinary Disks and Their Binary Hosts}",
      journal = {\apj},
         year = 2019,
        month = sep,
       volume = {883},
       number = {1},
          eid = {22},
        pages = {22},
          doi = {10.3847/1538-4357/ab287b},
archivePrefix = {arXiv},
       eprint = {1906.03269},
 primaryClass = {astro-ph.EP},
       adsurl = {https://ui.adsabs.harvard.edu/abs/2019ApJ...883...22C}
}

@ARTICLE{Duchene.Multiplicity.2013,
       author = {{Duch{\^e}ne}, Gaspard and {Kraus}, Adam},
        title = "{Stellar Multiplicity}",
      journal = {\araa},
         year = 2013,
        month = aug,
       volume = {51},
       number = {1},
        pages = {269-310},
          doi = {10.1146/annurev-astro-081710-102602},
archivePrefix = {arXiv},
       eprint = {1303.3028},
 primaryClass = {astro-ph.SR},
       adsurl = {https://ui.adsabs.harvard.edu/abs/2013ARA&A..51..269D}
}

@ARTICLE{Cuello.CBD.Review,
       author = {{Cuello}, Nicol{\'a}s and {Alaguero}, Antoine and {Poblete}, Pedro P.},
        title = "{Circumstellar and Circumbinary Discs in Multiple Stellar Systems}",
      journal = {Symmetry},
         year = 2025,
        month = feb,
       volume = {17},
       number = {3},
          eid = {344},
        pages = {344},
          doi = {10.3390/sym17030344},
archivePrefix = {arXiv},
       eprint = {2501.19249},
 primaryClass = {astro-ph.EP},
       adsurl = {https://ui.adsabs.harvard.edu/abs/2025Symm...17..344C}
}

@ARTICLE{Artymowicz.1994.CBDCavSize,
       author = {{Artymowicz}, Pawel and {Lubow}, Stephen H.},
        title = "{Dynamics of Binary-Disk Interaction. I. Resonances and Disk Gap Sizes}",
      journal = {\apj},
         year = 1994,
        month = feb,
       volume = {421},
        pages = {651},
          doi = {10.1086/173679},
       adsurl = {https://ui.adsabs.harvard.edu/abs/1994ApJ...421..651A}
}

@ARTICLE{Alexander.2012.BinaryAccretion,
       author = {{Alexander}, Richard},
        title = "{The Dispersal of Protoplanetary Disks around Binary Stars}",
      journal = {\apjl},
         year = 2012,
        month = oct,
       volume = {757},
       number = {2},
          eid = {L29},
        pages = {L29},
          doi = {10.1088/2041-8205/757/2/L29},
archivePrefix = {arXiv},
       eprint = {1209.0779},
 primaryClass = {astro-ph.EP},
       adsurl = {https://ui.adsabs.harvard.edu/abs/2012ApJ...757L..29A}
}

@ARTICLE{Ronco.2021.CBDLifeTime,
       author = {{Ronco}, Mar{\'\i}a Paula and {Guilera}, Octavio M. and {Cuadra}, Jorge and {Miller Bertolami}, Marcelo M. and {Cuello}, Nicol{\'a}s and {Fontecilla}, Camilo and {Poblete}, Pedro and {Bayo}, Amelia},
        title = "{Long Live the Disk: Lifetimes of Protoplanetary Disks in Hierarchical Triple-star Systems and a Possible Explanation for HD 98800 B}",
      journal = {\apj},
         year = 2021,
        month = aug,
       volume = {916},
       number = {2},
          eid = {113},
        pages = {113},
          doi = {10.3847/1538-4357/ac0438},
archivePrefix = {arXiv},
       eprint = {2105.09410},
 primaryClass = {astro-ph.EP},
       adsurl = {https://ui.adsabs.harvard.edu/abs/2021ApJ...916..113R}
}

@ARTICLE{Lodato.2017.DiscEvolution,
       author = {{Lodato}, Giuseppe and {Scardoni}, Chiara E. and {Manara}, Carlo F. and {Testi}, Leonardo},
        title = "{Protoplanetary disc `isochrones' and the evolution of discs in the Ṁ-M$_{d}$ plane}",
      journal = {\mnras},
         year = 2017,
        month = dec,
       volume = {472},
       number = {4},
        pages = {4700-4706},
          doi = {10.1093/mnras/stx2273},
archivePrefix = {arXiv},
       eprint = {1708.09467},
 primaryClass = {astro-ph.SR},
       adsurl = {https://ui.adsabs.harvard.edu/abs/2017MNRAS.472.4700L}
}

@ARTICLE{Manara.Frasca.ChromNoise.2017,
       author = {{Manara}, C.~F. and {Frasca}, A. and {Alcal{\'a}}, J.~M. and {Natta}, A. and {Stelzer}, B. and {Testi}, L.},
        title = "{An extensive VLT/X-shooter library of photospheric templates of pre-main sequence stars}",
      journal = {\aap},
         year = 2017,
        month = sep,
       volume = {605},
          eid = {A86},
        pages = {A86},
          doi = {10.1051/0004-6361/201730807},
archivePrefix = {arXiv},
       eprint = {1705.10075},
 primaryClass = {astro-ph.SR},
       adsurl = {https://ui.adsabs.harvard.edu/abs/2017A&A...605A..86M}
}

@INPROCEEDINGS{Offner.Multiplicity,
       author = {{Offner}, S.~S.~R. and {Moe}, M. and {Kratter}, K.~M. and {Sadavoy}, S.~I. and {Jensen}, E.~L.~N. and {Tobin}, J.~J.},
        title = "{The Origin and Evolution of Multiple Star Systems}",
    booktitle = {Protostars and Planets VII},
         year = 2023,
       editor = {{Inutsuka}, S. and {Aikawa}, Y. and {Muto}, T. and {Tomida}, K. and {Tamura}, M.},
       series = {Astronomical Society of the Pacific Conference Series},
       volume = {534},
        month = jul,
        pages = {275},
          doi = {10.48550/arXiv.2203.10066},
archivePrefix = {arXiv},
       eprint = {2203.10066},
 primaryClass = {astro-ph.SR},
       adsurl = {https://ui.adsabs.harvard.edu/abs/2023ASPC..534..275O}
}

@ARTICLE{Swastik.2026.WISPIT2,
       author = {{Swastik}, C. and {Bestha}, M. and {Sonith}, L.~S. and {Facchini}, S. and {Sivarani}, T. and {Banyal}, R.~K. and {Wahhaj}, Z. and {Choudhary}, A. and {Gopinathan}, M. and {Saraf}, P. and {Mallick}, A. and {Navaneeth}, M.~P. and {Biswas}, S. and {Khushbu}, K.},
        title = "{A Quiet Host in an Active Planet-Forming Disk: Optical Spectroscopy of WISPIT}",
      journal = {\aap},
         year = 2026,
        month = jul,
        volume = {712},
        eid = {L13},
        pages = {L13},
          doi = {10.48550/arXiv.2607.20649},
archivePrefix = {arXiv},
       eprint = {2607.20649},
 primaryClass = {astro-ph.SR},
       adsurl = {https://ui.adsabs.harvard.edu/abs/2026arXiv260720649S}
}

\begin{appendix}
\nolinenumbers

\onecolumn
\section{Observational information}

\subsection{Observations and data reduction}
\label{app:obs}
The two epochs of X-shooter spectra were observed within the DDT program 116.2ASZ (PI B\"{u}rgy). The first spectrum was taken on April 16, 2026 at airmass-corrected seeing $\sim1.3"$, above the requested $<1"$. Therefore, the observations were repeated on May 23, 2026 and carried out at airmass-corrected seeing $\sim1"$, as requested. During the first observations sky conditions were "Clear", while in the second epoch "Photometric". 

Both spectra were obtained with a narrow slit for the highest available spectral resolution and a wide slit for absolute flux calibration. The narrow slit observations were obtained using 0.5", 0.4", and 0.4" wide slits for the UVB, VIS, and NIR arms, respectively. Observations were performed in ABBA nodding mode to remove background sky emission. The wide slit observations were performed in stare mode using the 5" slit at the beginning of the observations. The corresponding reported spectral resolution of this observing setup is R $\sim$ 9700, 18400, and 11600 for the three arms, respectively.

The X-shooter data was reduced using the X-shooter data reduction pipeline \citep[v. 3.8.3;][]{XSHOOTER.Pipeline.Modigliani2010}, within the EsoReflex workflow \citep[v. 2.11.5;][]{EsoReflex.Freudling2013}, at default settings with the exception of the 1D spectral extraction window which was adapted to exclude a second star located within the slit for one of the nodding positions. Subsequently, the reduced spectra of the VIS and NIR arm were corrected for telluric absorption using Molecfit \citep[v. 4.4.4.82][]{Molecfit.Smette2015, Molecfit.Kausch2015}. The resulting spectra were re-scaled in flux to match the spectra taken with the wide slit in order to correct for flux losses. Finally, the spectra were individually corrected for the Earth's motion relative to the local standard of rest (LSRK) at the time of observation.

The FEROS observations were carried out under the DDT program 117.2AS7.001 (PI Benisty) between July 9 and July 13 2026, with an exposure time of 900~s each. The spectra were reduced using the \texttt{ceres} reduction pipeline\footnote{\url{https://github.com/rabrahm/ceres.git}} for echelle spectrographs \citep{Brahm.2017.Ceres.DataRed}. The reduced and continuum-normalised spectra were corrected for Earth's motion relative to LSRK at the time of observation.

\begin{table}[h]
    \caption{Log of X-shooter observations}
    \centering
    \begin{tabular}{c c c c c c c c}
    \hline\hline
        Date & MJD & Exp. Time & Slit Width & Airmass & Seeing & SNR & RV \\
        & (UVB - VIS - NIR) & (UVB - VIS - NIR) & & (end) & @ 705 nm & [km/s] \\
        \hline
        2026-04-16 & 61147.39 & 2 x 180s & 0.5 - 0.4 - 0.4 & 1.09 & 1.40" & 47.6 & $-16.72 \pm 0.4$ \\ 
        & & 2 x 90s & \\
        & & 2 x 4 x 50s & \\
        \hline
        
        2026-05-23 & 61184.29 & 2 x 180s & 0.5 - 0.4 - 0.4 & 1.08 & 1.01" & 49.6  & $24.51 \pm 0.7$ \\ 
        & & 2 x 90s & \\
        & & 2 x 4 x 50s & \\
        \hline
    \end{tabular}
    \label{tab:XS_ObsLog}
    \tablefoot{Observational set up used to take the X-shooter data. Additionally, we report the RV measurements of WISPIT~2 in LSRK, derived via cross-correlation with a photospheric template.}
\end{table}

\begin{table}[h]
\caption{Log of FEROS observations}
    \centering
    \begin{tabular}{c c c c c c c}
    \hline\hline
        Date & MJD & Exp. Time & Airmass & Seeing & RV [km/s] & Phase ($\phi$)\\
        \hline
        2026-07-09 & 61231.25 & 900.0932s & 1.12 & 1.04 & $25.15 \pm 0.10$ &  0.8576 \\ 
        2026-07-10 & 61232.25 & 900.0463s & 1.13 & 1.63 & $27.91 \pm 0.09$ & 0.0656 \\ 
        2026-07-11 & 61233.26 & 900.0464s & 1.15 & 1.07 & $-0.44 \pm 0.09$ & 0.2739 \\ 
        2026-07-12 & 61234.25 & 900.0460s & 1.13 & 0.98 & $-18.45 \pm 0.08$ & 0.4798 \\ 
        2026-07-13 & 61235.25 & 900.0464s & 1.16 & 0.83 & $-0.84 \pm 0.11$ &0.6883 \\ 
        \hline
    \end{tabular}
    \label{tab:FEROS_ObsLog}
    \tablefoot{Observational set up used to take the FEROS data. Additionally, we report the RV measurements of WISPIT~2 in LSRK, derived via cross-correlation with a photospheric template.}
\end{table}

\newpage
\section{Spectra}

\begin{figure}[ht!]
    \centering
    \includegraphics[width=0.7\linewidth]{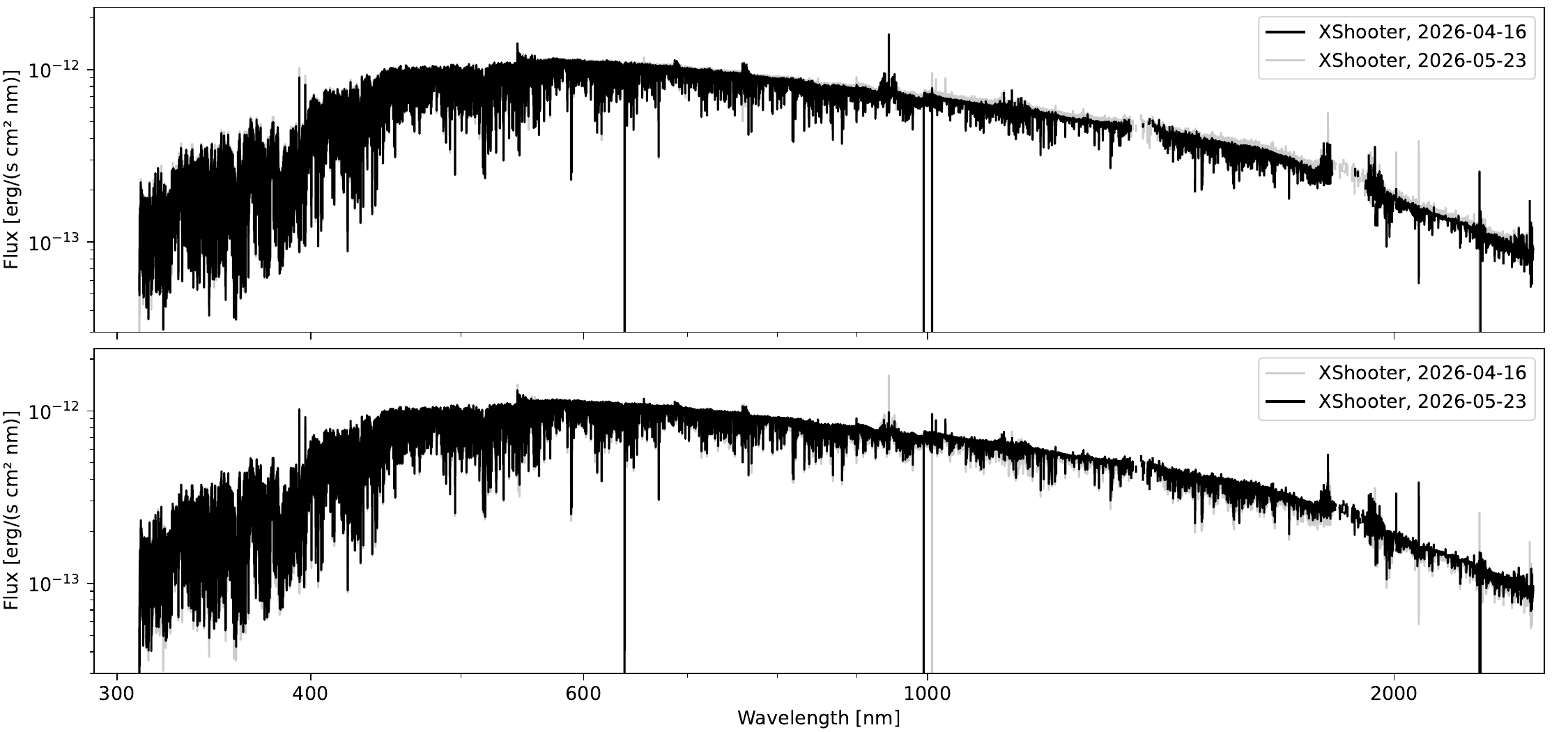}
    \caption{Overview of the two X-shooter spectra obtained. For comparison, the respective other epoch is plotted in grey.}
    \label{fig:XSSpec}
\end{figure}

\begin{figure}[ht]
    \centering
    \includegraphics[width=0.7\linewidth]{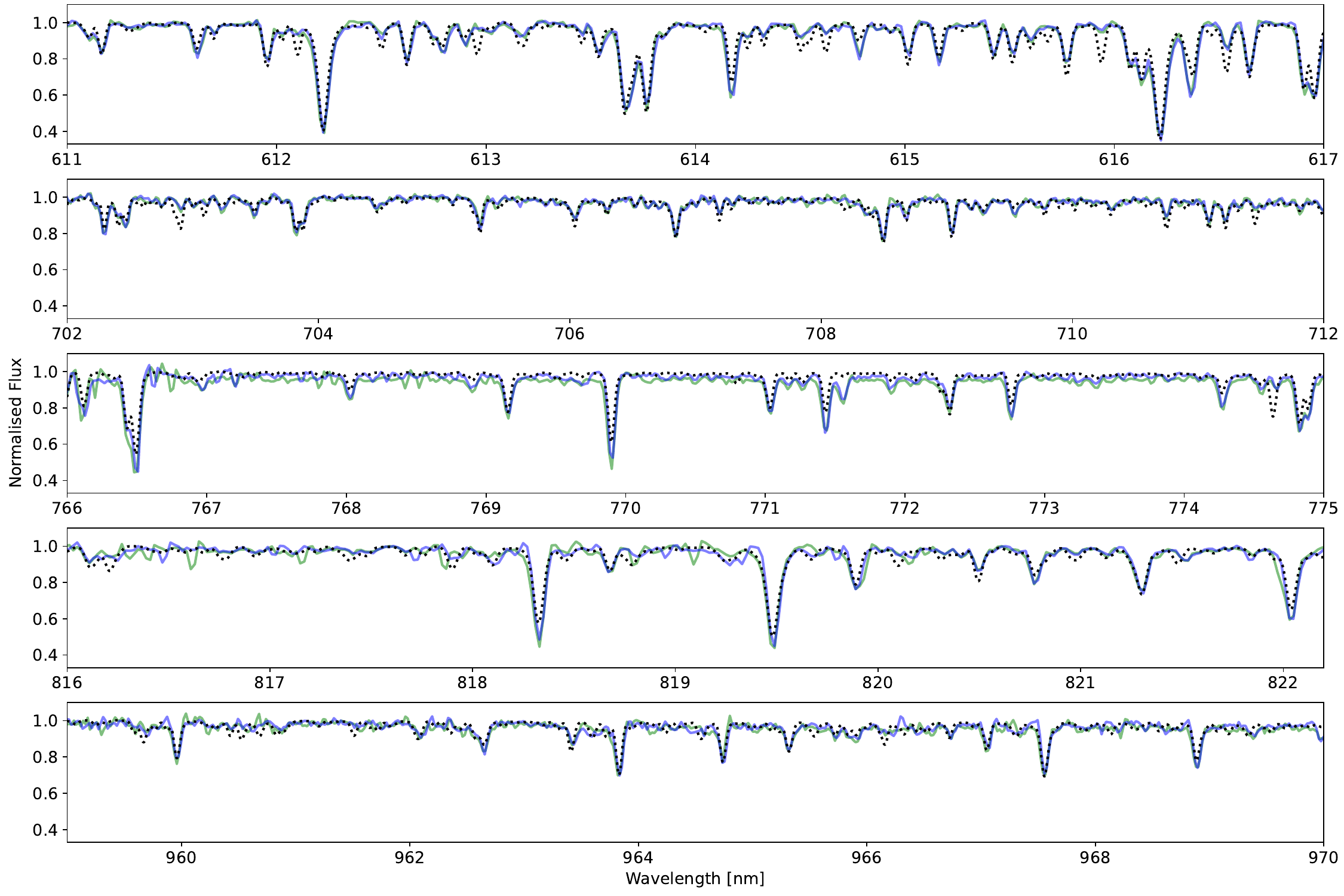}
    \caption{Demonstration of the fitted PHOENIX model. Shown are the two X-shooter epochs in blue and green in the background, as well as the best fit PHOENIX model in dotted black.}
    \label{fig:PhotFit}
\end{figure}

\begin{table*}[h]
    \caption{Spectral windows used for photospheric analysis}
    \centering
    \begin{tabular}{c c | c c || c c | c c}
    \hline\hline
        \multicolumn{4}{c|}{X-shooter} & \multicolumn{4}{c}{FEROS}\\
        Window & Arm & Window & Order & Window & Order & Window & Order\\
        \hline 
        440 - 459 nm & UVB & 816 - 822 nm & VIS & 517 - 533 nm & 9 & 603 - 622 nm & 3\\
        512 - 522 nm & UVB & 959 - 980 nm & VIS & 533 - 545 nm & 8 & 617 - 626 nm & 2\\
        534 - 550 nm & VIS & & & 545 - 557 nm & 7 & 634 - 638 nm  & 2\\
        605 - 627 nm & VIS & & & 557 - 573 nm & 6 & 638 - 645 nm & 1\\
        702 - 712 nm & VIS & & & 573 - 588 nm & 5 & 650 - 653 nm & 1\\
        760 - 772 nm & VIS & & & 588 - 603 nm & 4 & &  \\
        \hline
    \end{tabular}
    \tablefoot{List of spectral windows used for the photospheric fits of the X-shooter spectra, as well as the cross-correlation of the FEROS spectra.}
    \label{tab:SpecWindows}
\end{table*}

\begin{figure*}
    \centering
    \includegraphics[width=0.8\linewidth]{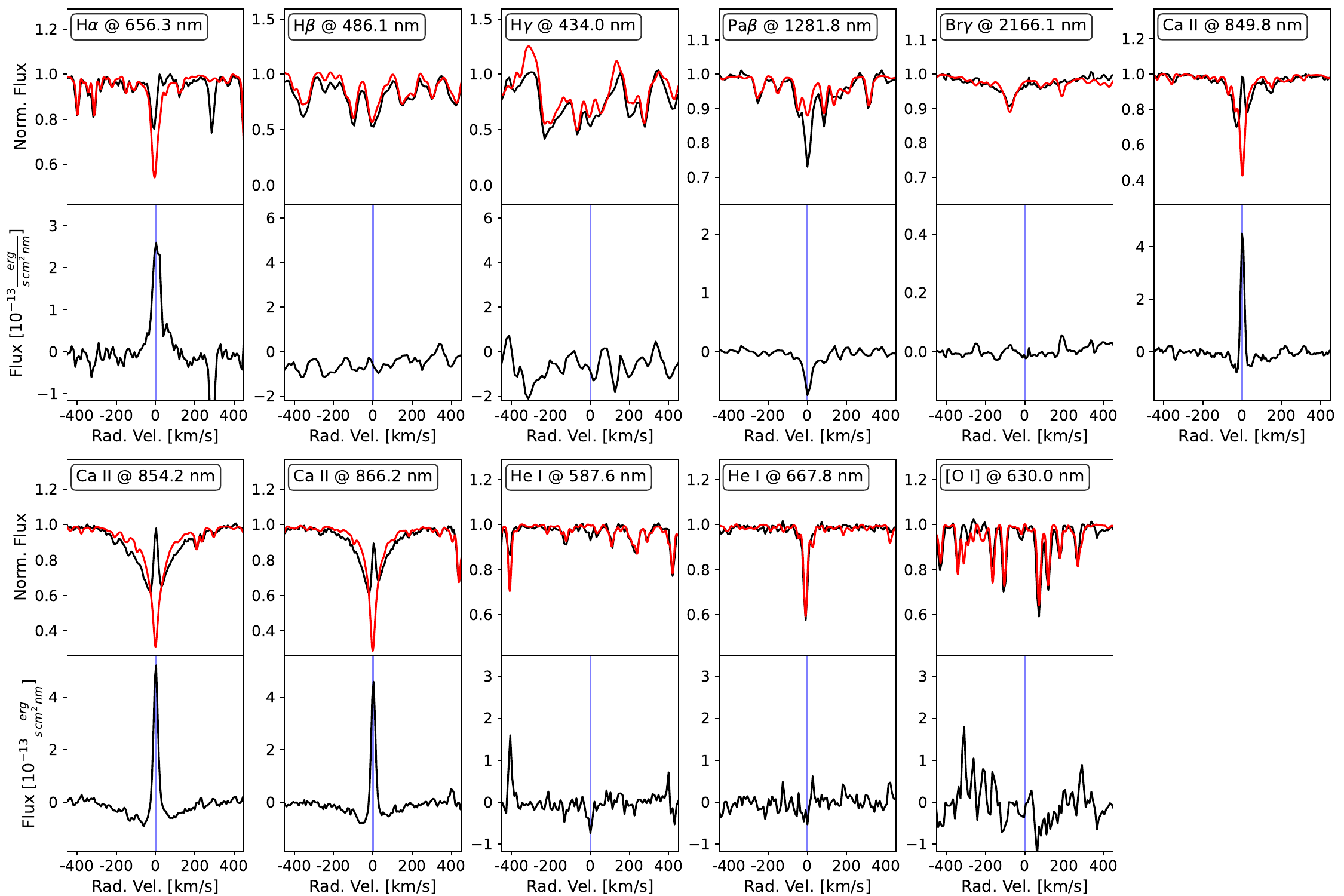}
    \caption{Gallery of lines traced with X-shooter on 2026-04-16. The top row shows the normalised X-shooter spectrum, shifted by $v_{r} = -24.29$ km s$^{-1}$ with respect to the disc systemic velocity (black), and the best fit PHOENIX stellar model convolved to the spectral resolution of the observations (red). The bottom row shows the residual flux.}
    \label{fig:LineGallery_ep1}
\end{figure*}

\begin{figure*}
    \centering
    \includegraphics[width=0.8\linewidth]{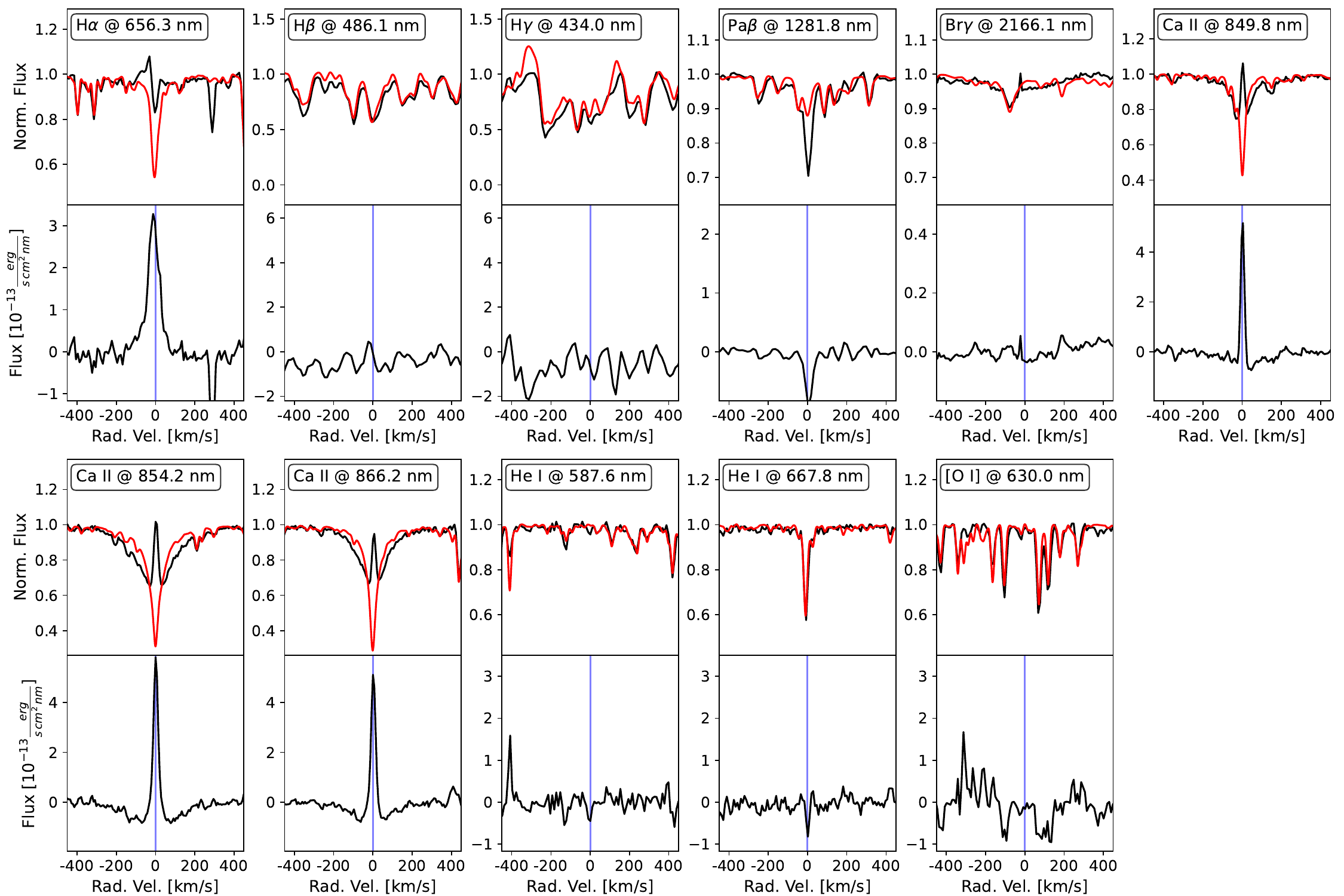}
    \caption{Gallery of lines traced with X-shooter on 2026-05-23. The top row shows the normalised X-shooter spectrum, shifted by $v_{r} = +20.11$ km s$^{-1}$ with respect to the disc systemic velocity (black), and the best fit PHOENIX stellar model convolved to the spectral resolution of the observations (red). The bottom row shows the residual flux.}
    \label{fig:LineGallery_ep2}
\end{figure*}

\begin{table*}[h]
    \caption{X-shooter line detections}
    \centering
    \begin{tabular}{c c c c c c c}
    \hline\hline
        Line & $\lambda_0$ & $\Delta v_0$ & FWHM & $F_{\rm line}$ & $L_{\rm line}$ & $\log{L_{\rm acc}/L_\odot}$ \\
         & (nm) & (km s$^{-1}$) & (km s$^{-1}$) & ($10^{-14}$ erg s$^{-1}$ cm$^{-2}$) & ($10^{-5} L_\odot$) &  \\
        \hline 
        \multicolumn{7}{c}{2026-04-16} \\
        \hline
        H$\alpha$ & $656.29$ & $5.72$ & $51.0$ & $3.44$ & $1.90$ & $-3.63$ \\
        Ca II & $849.80$ & $1.47$ & $19.9$ & $2.77$ & $1.53$ & $-1.63$ \\
        Ca II & $854.21$ & $1.05$ & $28.8$ & $5.11$ & $2.83$ & $-1.43$ \\
        Ca II & $866.21$ & $2.28$ & $25.6$ & $4.04$ & $2.24$ & $-1.33$ \\
        \hline 
        \multicolumn{7}{c}{2026-05-23} \\
        \hline
        H$\alpha$ & $656.29$ & $-10.14$ & $68.5$ & $5.27$ & $2.91$ & $-3.39$ \\
        Ca II & $849.80$ & $1.70$ & $20.7$ & $3.46$ & $1.91$ & $-1.53$ \\
        Ca II & $854.21$ & $1.21$ & $30.7$ & $6.22$ & $3.44$ & $-1.34$ \\
        Ca II & $866.21$ & $2.16$ & $27.9$ & $4.96$ & $2.74$ & $-1.24$ \\
        \hline
    \end{tabular}
    \tablefoot{Overview of the detected lines. For both epochs, we report the observed centroid shift $\Delta v_0$, the line FWHM, and the integrated line flux for the H$\alpha$ line, as well as the CaIRT.}
    \label{tab:LineDets}
\end{table*}

\begin{table*}[h]
    \caption{H$\alpha$ variability}
    \centering
    \begin{tabular}{c c | c c}
    \hline \hline
        Night & MJD & EW [nm] & Phase ($\Phi$) \\
        \hline
        2026-04-16 & 61147.39 & 0.035 & 0.1149 \\
        2026-05-23 & 61184.29 & 0.052 & 0.4584 \\
        2026-07-09 & 61231.25 & 0.022 & 0.8576 \\
        2026-07-10 & 61232.25 & 0.014 & 0.0656 \\
        2026-07-11 & 61233.26 & 0.024 & 0.2739 \\
        2026-07-12 & 61234.25 & 0.026 & 0.4798 \\
        2026-07-13 & 61235.25 & 0.027 & 0.6883 \\
        \hline
    \end{tabular}
    \tablefoot{Equivalent width of the H$\alpha$ emission line throughout all available epochs.}
    \label{tab:HalphaEW}
\end{table*}

\begin{figure*}
    \centering
    \includegraphics[width=0.35\linewidth]{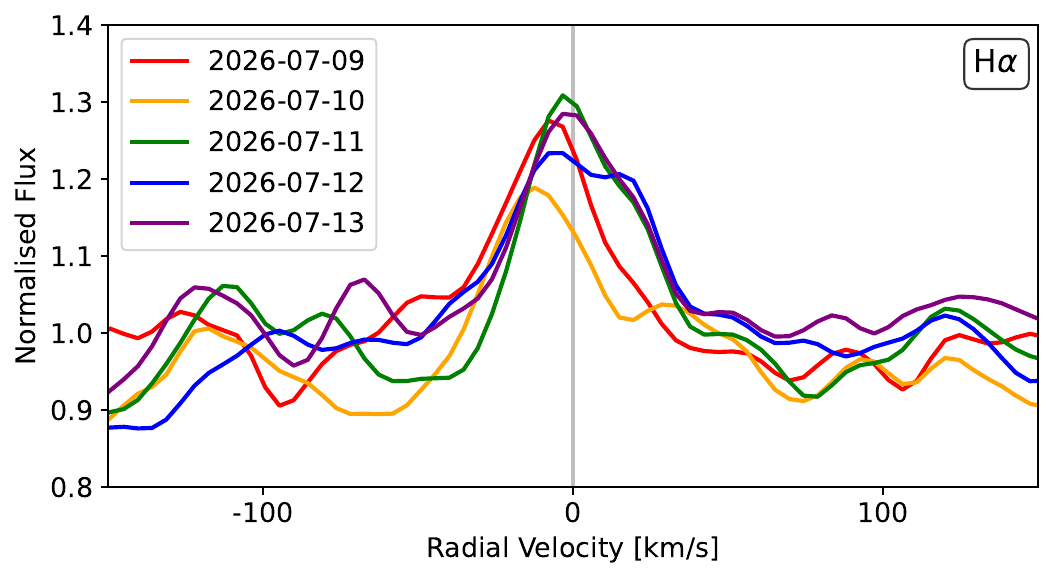}
    \caption{Comparison of H$\alpha$ line profiles across the FEROS epochs. The spectra are smoothed to the X-shooter instrumental resolution to improve SNR.}
    \label{fig:HalphaVariab}
\end{figure*}

\begin{figure*}
    \centering
    \includegraphics[width=0.4\linewidth]{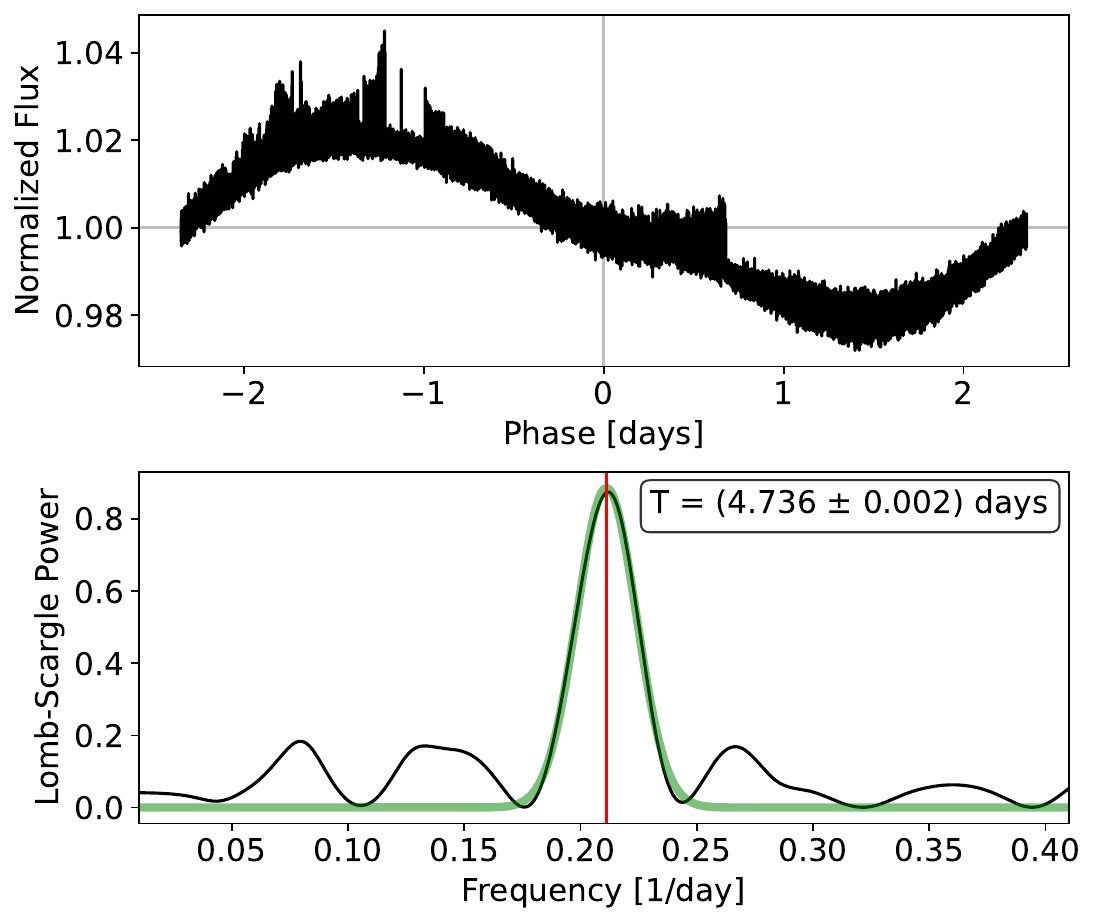}
    \caption{TESS light curve folded at the period of $P \sim 4.74$ days (top), and periodogram (bottom), with fitted Gaussian (green).}
    \label{fig:TESS}
\end{figure*}

\FloatBarrier
\twocolumn
\section{Radial velocity analysis}

\begin{figure}[h!]
    \centering
    \includegraphics[width=\linewidth]{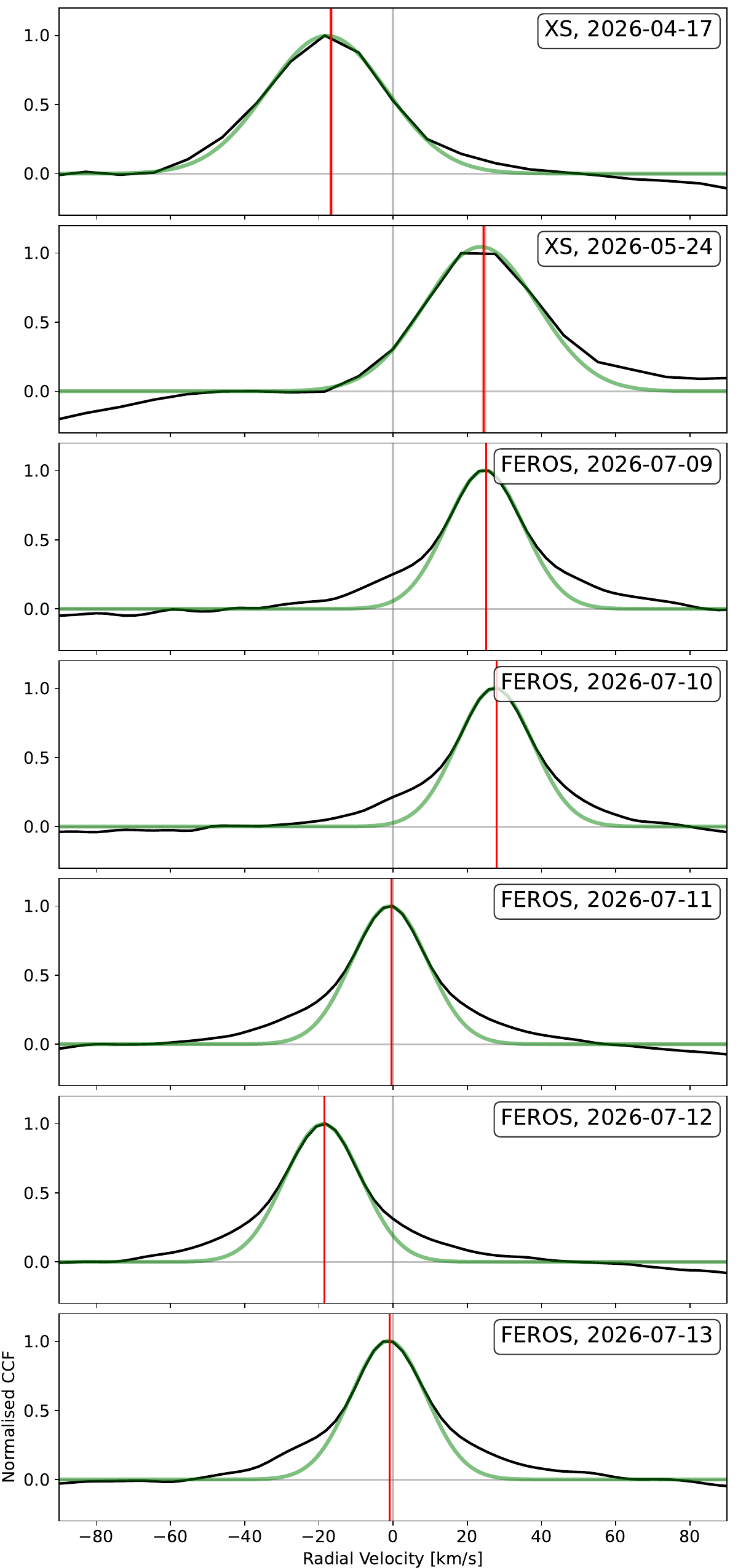}
    \caption{Normalised cross correlation functions for all available epochs. We fit a Gaussian to the core-region around the cross correlation peak, which is shown in green. The derived RV centroid is over-plotted in red.}
    \label{fig:CCF}
\end{figure}

\FloatBarrier 

The fit of the RV curve described in Sect.\ref{sec:RV}, which corresponds to the solution of P$\sim 4.82$. The distributions are single-peaked, with Gaussian profiles. The off-diagonal 2D plots represent the bivariate distributions between each pair of parameters, which provide an immediate estimate of their correlation. 

\begin{figure}[h!]
    \centering
    \includegraphics[width=\linewidth]{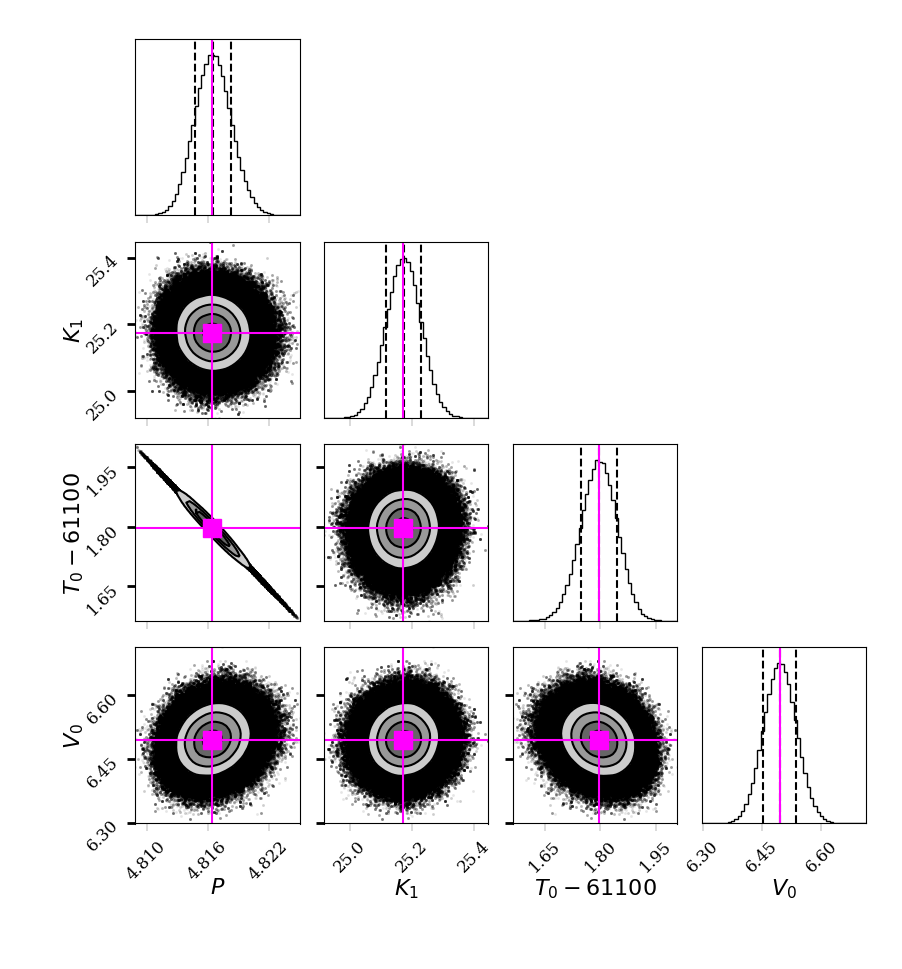}%
    \vspace{-5pt}
    \caption{Corner plot of the orbital solution for WISPIT~2 adapted in this work. The vertical dashed lines indicate the 16th, 50th and 84th percentile. The vertical magenta lines correspond to the best likelihood estimates of each distribution. }
    \label{fig:MCMC_4_82_days}
\end{figure}

\begin{figure}[h!]
    \centering
    \includegraphics[width=\linewidth]{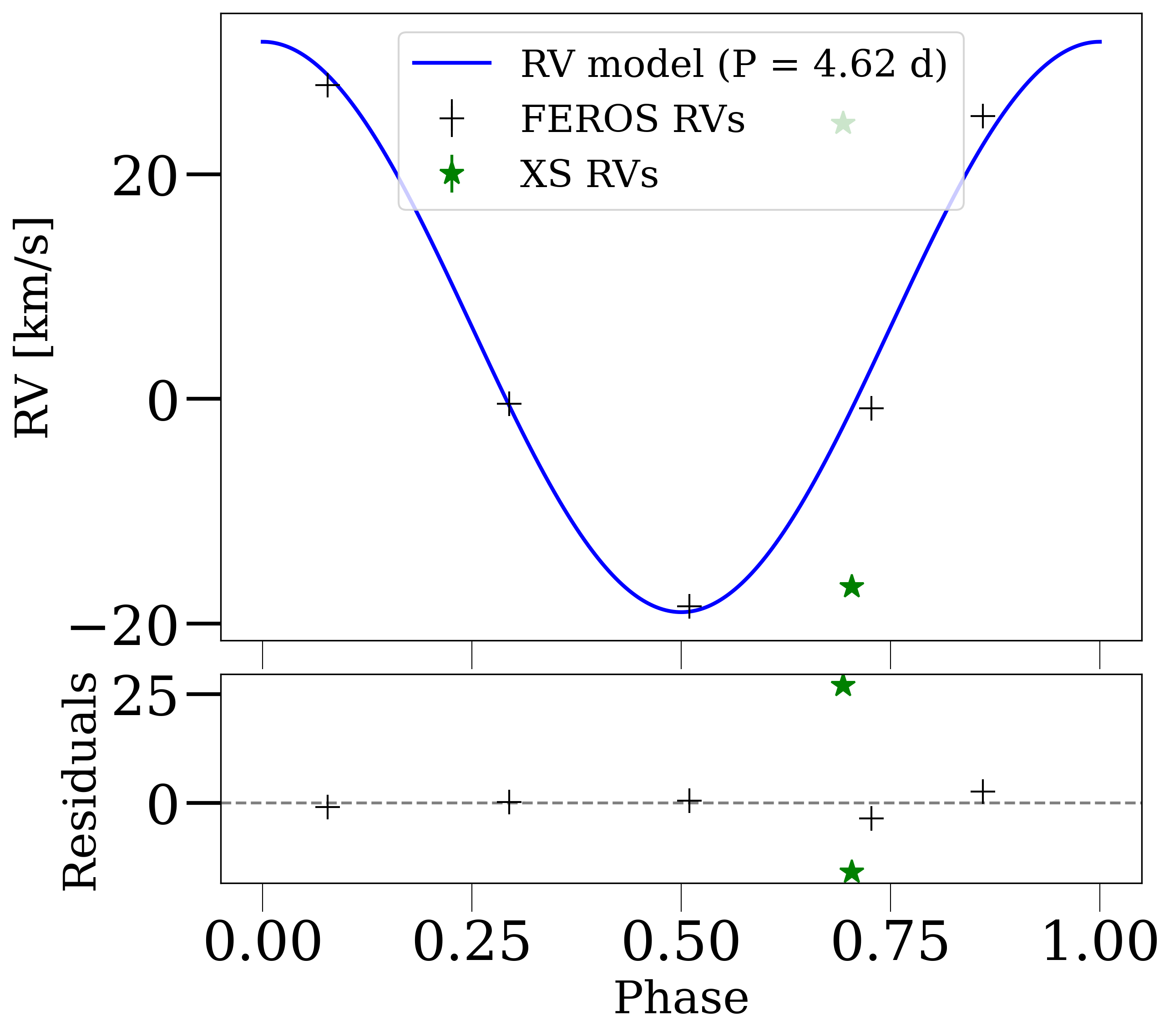}
    \vspace{-5pt}
    \caption{Phase-folded solution of WISPIT~2 with P$\sim4.62$ days. The blue line fits the RV data taken with VLT/X-shooter in green and FEROS in black. Below, the residuals of the fitted model in phase.} 
    \label{fig:5_15_fit}
\end{figure}

\end{appendix}
\end{document}